\documentclass[%
nobibnotes,
nofootinbib,
longbibliography,
reprint,
twocolumn,
 amsmath,amssymb,
 aps,
 prd,
floatfix,
]{revtex4-2} 

\usepackage{graphicx}
\usepackage{dcolumn}
\usepackage{bm}

\usepackage{amsmath}
\usepackage{adjustbox}
\usepackage{booktabs}
\usepackage{sidecap}
\usepackage{floatrow}
\usepackage{float}
\usepackage{tabularx}

\usepackage{siunitx}

\usepackage[colorlinks=true,allcolors=blue]{hyperref}
\usepackage{cleveref}

\newfloatcommand{capbtabbox}{table}[][\FBwidth]
\newcommand{\sibyll}[1]{{\sc Sibyll#1}}

\newcommand{\mceq}{{\sc MCEq}}
\newcommand{\daemonflux}{{\sc daemonflux}}
\newcommand{\ddm}{DDM}
\newcommand{\rmd}{{\mathrm d}}
\newcommand{\xl}{x_{\rm Lab}}

\newcommand{\muratio}{$R_{\mu^{+}/\mu^-}$}
\newcommand{\muflux}{$\Phi_\mu$}
\newcommand{\degrees}{$^\circ$}

\newcommand{\pions}{$\pi^{\pm}$}
\newcommand{\kaons}{$K^{\pm}$}

\newcommand{\lepip}{{\pi^+_\mathrm{31G}}}
\newcommand{\lepim}{{\pi^-_\mathrm{31G}}}

\newcommand{\hepip}{{\pi^+_\mathrm{158G}}}
\newcommand{\hepim}{{\pi^-_\mathrm{158G}}}
\newcommand{\heKp}{{K^+_\mathrm{158G}}}

\newcommand{\vheIpip}{{^\bigstar \pi^+_\mathrm{20T}}}
\newcommand{\vheIpim}{{^\bigstar \pi^-_\mathrm{20T}}}

\newcommand{\GSFI}{{\mathrm{GSF}_\mathrm{1}}}

\newcommand{\GSFV}{{\mathrm{GSF}_\mathrm{5}}}
\newcommand{\GSFVI}{{\mathrm{GSF}_\mathrm{6}}}

\begin{document}

\preprint{APS/123-QED}

\title{{\sc daemonflux}: DAta-drivEn MuOn-calibrated atmospheric Neutrino Flux}

\author{Juan Pablo Ya\~nez}
\email{j.p.yanez@ualberta.ca}
\affiliation{Dept.\ of Physics, University of Alberta\\
Edmonton, Alberta, Canada T6G 2E1}%
\author{Anatoli Fedynitch}
\email{anatoli@gate.sinica.edu.tw}
\affiliation{Institute of Physics, Academia Sinica, Taipei City, 11529, Taiwan}
\affiliation{Institute for Cosmic Ray Research, the University of Tokyo,
  5-1-5 Kashiwa-no-ha, Kashiwa, Chiba 277-8582, Japan}

\date{\today}

\begin{abstract}
In this paper, we present a refined calculation of the atmospheric neutrino flux spanning from GeV to PeV energies. Our method, Daemonflux, utilizes data-driven inputs and incorporates adjustable parameters to take their uncertainties into account. By optimizing these parameters using a combination of muon data and constraints from fixed-target experiments, we achieve uncertainties in the calculated neutrino fluxes of less than 10\% up to 1 TeV, with neutrino ratios constrained to below 10\%. Our model performs particularly well at energies below 100~GeV, where the smallest errors are obtained. We make our model available as a software package that provides access to predictions of fluxes, ratios, and errors, including the covariance matrix obtained from the fit.

\end{abstract}

\maketitle


\section{Introduction}

Atmospheric neutrinos are a product of the hadronic component of particle showers triggered by cosmic rays interacting with Earth's atmosphere. The majority of these neutrinos are produced as a result of the decay of $\pi$ and $K$ mesons, which also give rise to muons. Some of these muons will undergo decay in flight, producing additional neutrinos in the process. For a comprehensive overview of the topic, see Ref.~\cite{Gaisser:2016uoy}. The atmospheric neutrino flux encompasses a broad energy range, from MeV to hundreds of TeV, and due to their small cross section, the Earth is effectively transparent to neutrinos up to their highest energies, making it possible to detect them from all directions in the sky.

Atmospheric neutrinos are both a tool for discovery, as demonstrated by their role in uncovering neutrino oscillations~\cite{Super-Kamiokande:1998kpq, MACRO:1998ckv, Soudan-2:1999jbo} and searches for new physics~\cite{Super-Kamiokande:2011dam,IceCube:2017ivd,ANTARES:2018rtf, IceCube:2020tka, IceCubeCollaboration:2021euf,ANTARES:2021crm,IceCube:2022ubv}, as well as the main background for the emerging field of neutrino astronomy~\cite{IceCube:2018cha,IceCube:2018dnn, IceCube:2020nig,IceCube:2022der}. Their understanding is central to the interpretation of data from operational neutrino observatories like Super-Kamiokande, IceCube, KM3NeT and Baikal-GVD, and future projects such as Hyper-Kamiokande, IceCube-Gen2, and P-ONE, to mention a few. However, despite their relevance, existing calculations from neutrino fluxes have significant uncertainties that considerably limit the precision of measurements conducted by these experiments.

One major source of uncertainty for the computation of neutrino fluxes comes from the limited knowledge of the primary cosmic rays that initiate the atmospheric showers~\cite{Evans:2016obt}. Historically the primary cosmic ray flux has been modeled as a collection of multiple nuclei that follow a power-law spectrum, fitting the scaling factor and spectral index to data from multiple experiments that were not always in agreement~\cite{Gaisser:2002jj}. The other main uncertainty in flux calculations comes from the phase space of hadronic interactions that is relevant for the observed neutrinos, namely light meson production at very small scattering angles~\cite{Barr:2006it}. This region falls under the domain of non-perturbative QCD, making it unfeasible to compute neutrino yields using first principles. Phenomenological models of hadronic interactions, such as \sibyll{-2.3d} \cite{Riehn:2019jet,Fedynitch:2018cbl}, are commonly used as an alternative. However, accurately modeling forward particle distributions in the relevant phase space requires comparing these models to data obtained from fixed-target experiments, which are typically not conducted at TeV energies and do not cover sufficient phase space.

One way to minimize these uncertainties is to utilize the correlation between the flux of atmospheric neutrinos and that of cosmic muons, which originate from the same decay of charged mesons and are comparatively easier to detect and characterize. The relationship between muons and neutrinos was first explored by Honda et al.~\cite{Honda:2006qj}, where data from two muon spectrometers was used to modify the meson yields predicted by a hadronic interaction model to match the muon flux data and charge ratio $\Phi_{\mu^+}/\Phi_{\mu^-}$. The muon-calibrated flux from Honda et al. has been successfully used to interpret neutrino data from experiments such as Super-Kamiokande (e.g.~\cite{Super-Kamiokande:2019gzr}) and IceCube (e.g.~\cite{IceCubeCollaboration:2021euf}). However, this approach had several limitations, as it was only applied to two data sets, did not account for uncertainties associated with cosmic rays, and lacked transparency in its methodology for users.

In this work we exploit the muon-neutrino relationship, extending it to include flux and muon charge ratio measurements from multiple experiments. These measurements are used as calibration data to modify the fluxes computed with the Global Spline Fit~\cite{Dembinski:2017zsh} and the Data Driven Model~\cite{Fedynitch:2022vty}, both in cosmic ray spectra and particle yields of the atmospheric showers, respectively. The result is \daemonflux{}\footnote{https://github.com/mceq-project/daemonflux} \cite{daemonflux-070}, a \textit{calibrated} atmospheric neutrino flux with a covariance matrix of well defined uncertainties.

\section{Calibration of neutrino fluxes}

The calibration of atmospheric neutrino fluxes is achieved by using muon flux and charge ratio observations to refine the uncertainty inputs of the calculation, specifically the cosmic ray spectra and the production of mesons in atmospheric showers. In this section, we provide a comprehensive overview of the calculation method, followed by a detailed explanation of the quantification of uncertainties.

\subsection{Evaluation of lepton fluxes using MCEq}

For this work we compute inclusive atmospheric lepton fluxes at energies above a few GeV using one-dimensional coupled cascade equations. Different solution techniques have been developed over the past decades (e.g., \cite{Gaisser:2016uoy,Gaisser:2002jj,Honda:2006qj,Barr:2004br,Bugaev:1998bi,Fedynitch:2012fs}) but at higher energies numerical (iterative) solutions are preferred due the advantage in speed at similar or higher level of detail compared to Monte Carlo calculations. To solve the cascade equations in this work we employ the code \mceq{} 1.4\footnote{https://github.com/mceq-project/MCEq}, which numerically solves for the evolution of particle densities as they propagate through a gaseous or dense medium.

The physics of \mceq{} has been described in more detail in Refs.~\cite{Fedynitch:2015zma, Fedynitch:2018cbl, Fedynitch:2022vty}, and its output has been compared to lepton flux and underground data in Refs.~\cite{Fedynitch:2022vty,Fedynitch:2021ima}, and to the CORSIKA 7 and 8 air-shower simulators in Ref.~\cite{CORSIKA8:2021gjr}. We forward the reader to these references for the technical details. 

\mceq{} technically covers an energy range from a few tens of MeV up to ZeV, but the relevant range for this work is $\sim\SIrange[range-phrase=-,range-units=single]{5}{10000}{GeV}$, motivated by the availability of surface muon flux data . Within this interval the muon production is governed by the production of charged pions and kaons in the projectile fragmentation region (secondary particles carrying a large momentum fraction of the projectile). Outside of this energy range other production channels become relevant such as the production of prompt muons at higher energies and, at lower energies, the geomagnetic effects, or the shift of the average muon production height due to variations of baryon production and inelasticity. Apart from the decay kinematics and branching fractions, the neutrino fluxes are described by the same dominant channels with the addition of muon decay, which dominates the lower-energy electron neutrino fluxes and has to be modeled including the muon polarization \cite{Lipari:1993hd}.

Since we are interested in percent-level precision, we do not apply any further simplifications to the cascade calculations. \mceq{} needs a complete set of realistic physical models and several choices are required to obtain the flux predictions:
\begin{enumerate}
    \item a cosmic ray nucleon flux model,
    \item a hadronic interaction model that predicts, differential particle yields as function of projectile and secondary particle energy,
    \item a (global) density profile of the atmosphere evaluated at
        \begin{enumerate}
            \item the locations where the muon flux measurements have been performed,
            \item and the locations for which the calibrated fluxes are computed.
        \end{enumerate}
\end{enumerate}

To avoid repeated cascade equation evaluations by minimization routines, we tabulate fluxes for each of the experiments we use in our calibration, taking into account the specific conditions of each measurement. These conditions include the reported zenith angles, the altitude, the atmosphere (averaged over the duration of data-taking using NRLMSISE-00 \cite{Picone:2002go}), and the variable in which the spectrum is reported (momentum or energy). More details about the application of the tabulated flux database can be found in Sec.~\ref{sec:impl_cal_params}.

\subsection{The inclusive hadronic interaction model DDM}

Hadronic event generators used in cosmic ray physics aim to simultaneously describe as much data as possible to retain some predictive power toward the highest energies or to parts of the phase space without any data coverage. A recent one is the \sibyll{-2.3d} model, which has been crosschecked with inclusive lepton fluxes during its development, and it was found that it could not reach an ideal flux description due to the model's complexity and broad scope, attempting to describe air-shower and collider physics. 

In an event generator like \sibyll{} it is almost impossible to obtain errors on inclusive cross sections due to the physical correlations among the model parameters, which affect the description of data globally, and only very rarely correct a single or a few particular deficiencies. Ultimately, the attempt to derive flux uncertainties by propagating errors on internal model parameters will result in a less realistic estimate compared to a specialized model, which is geared toward a specific purpose \cite{Parsons:2011ad}. The Data-Driven hadronic interaction Model (\ddm{})~\cite{Fedynitch:2022vty} is such a model specifically designed for lepton flux calculations and comes with advantages for this particular work.

In a nutshell, \ddm{} is a collection of fits to a carefully selected set of inclusive particle production cross section measurements (particle yields) from fixed target accelerators. The details about the choice of data and the additional assumptions are discussed in high detail in Ref.~\cite{Fedynitch:2022vty}. A novelty of DDM compared to previous approaches, such as the Kimel-Mokhov model \cite{Kimel:1974sn}, is the parameterization of cross section uncertainties directly from data. There are, however, some additional uncertainties that can not be constrained by accelerator data alone and require modelling decisions. These are, in decreasing order of importance
\begin{enumerate}
    \item the energy dependence of the longitudinal shape of inclusive cross sections, the interpolation between measurements taken at different energies, and the extrapolation beyond fixed-target energies;
    \item the extrapolation into very forward phase space that technically can not be seen by current detector designs;
    \item the application of event generators to correct for missing yields, such as the absence of charged kaon measurements on nuclear targets at high-energy;
    \item the isospin symmetries that govern the ``mirroring'' of cross sections from proton to neutron projectiles (and separately for pion projectiles), which are also used to parameterize neutral kaons from charged kaon yields (K$^0_\text{S}$ + K$^0_\text{L}$ = K$^+$ + K$^-$);
    \item and the uncorrected measurement errors, such as feed-down, of the fixed-target data data itself (relevant for older and modern spectrometer measurements \cite{Fischer:2022oqp}).
\end{enumerate}
Isospin symmetry is approximately conserved at small scattering angles and for light flavor hadrons. The other assumptions remain uncertain and challenging to quantify with regards to error. As shown in \cite{Fedynitch:2022vty}, the predictions from event generators cannot be used as an unambiguous guide for, \textit{e.g.}~the energy dependence, since all tested ``semi-phenomenological'' models predict a different energy scaling of the spectrum weighted moments due to differences in their underlying theory. In this regard, the calibration performed here is capable of shedding more light on points 1.~and~2., and can provide indirect constraints on the energy dependence of yields and the extrapolation into the missing phase space.

The \ddm{} contains in total 11 inclusive single-differential cross sections parametrized from 31~GeV and 158~GeV proton--carbon and proton--proton fixed target data \cite{Abgrall:2015hmv, Alt:2006fr,Anticic:2010yg}, and 12 cross sections from $\pi^-$--carbon data taken by NA61 at 158~GeV and 350~GeV \cite{Prado:2018wsv}. For the calculation of inclusive lepton fluxes, the impact of differences in the modeling of pion interactions and in anti-baryon production are negligible. Therefore, only 10 of the 23 hadron production cross sections for $\pi^\pm$, $K^\pm$, p and n are relevant for the present work.

\subsection{The Global Spline Fit of cosmic ray fluxes}
The Global Spline Fit (GSF) model represents the measurements of cosmic ray fluxes \cite{Dembinski:2017zsh}. It merges the spectra of individual elements from direct space and high-altitude balloon measurements with indirect observations from air shower arrays, which measure mass groups. To address the difference, GSF combines the lower-energy spectra of elements into four mass groups: hydrogen group (H), helium group (He), oxygen group (Li-Ne), and iron group (Na-Ni). It uses four B-spline curves to interpolate between datasets in $\log_{10}{R}$ (where $R$ is the rigidity). The data used, including systematic uncertainties, are from carefully selected experiments. The largest uncertainties are energy scales, affecting both the flux normalization and energy measurement. The spline coefficients and energy scales are determined by fitting the composition and all-particle flux measurements simultaneously. When spectrometer data is available, the element spectra are fitted by their own spline in rigidity, then combined into a mass group for comparison with air shower data. The relative ratios within one mass group (e.g. Mg/Fe) are kept constant outside of the range of direct measurements, one of the few physical assumptions of the model.

Unlike other models, GSF does not impose bias through assumptions about functional shapes motivated by cosmic ray acceleration and transport theories. However, GSF requires many parameters ($\sim90$), most of which are not relevant for this study, which depends mostly on  the proton and helium groups dominating the cosmic ray nucleon flux \cite{Gaisser:2002jj}. In the following, we outline the steps taken to simplify the model by reducing the number of parameters.

\begin{figure}
    \includegraphics[width=\columnwidth]{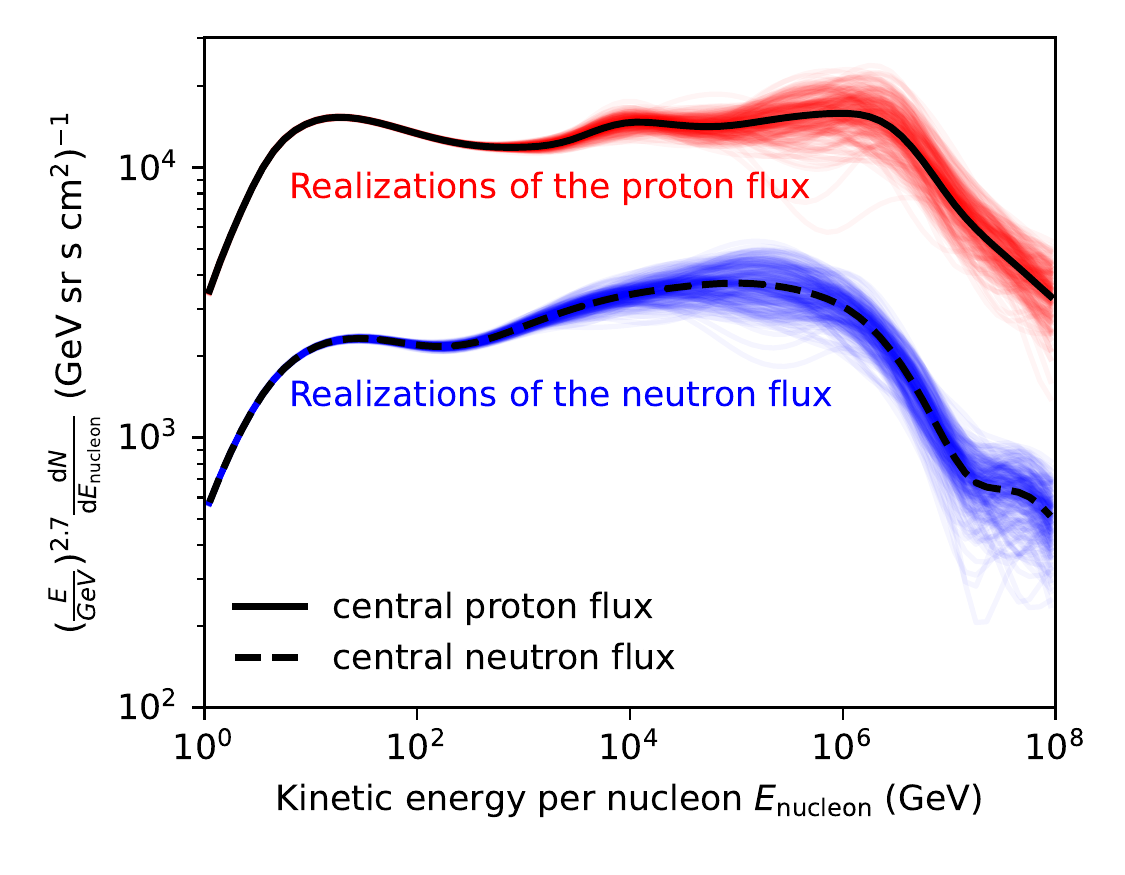}  
    \caption{Simulated variations of the proton and neutron fluxes based on the uncertainties of the GSF model. These realizations were generated using the covariance matrix of the model's parameters. At energies above 10 TeV, the model exhibits increased spread due to the limited precision of direct cosmic ray experiments and the absence of reliable indirect data that is sensitive to composition.\label{fig:GSF_samples}}
\end{figure}
\begin{figure}
    \includegraphics[width=\columnwidth]{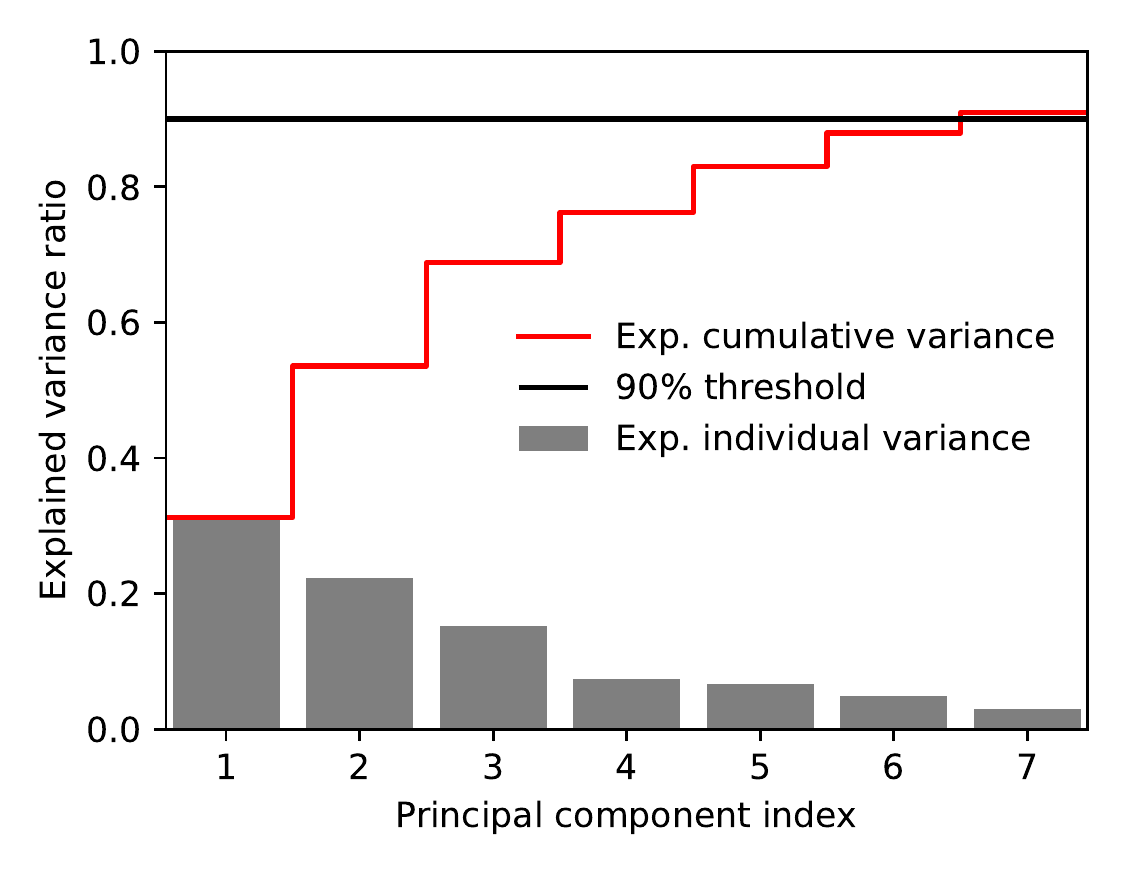}  
    \caption{Variance explained by the largest principal components (PCs). Almost 90\% of the total variance are retained using only 6 PCs (parameters). By using fewer PCs, the flux realizations would scatter less in the knee region because most of the variance stems from above 100 TeV. For other practical purposes, GSF-PCA realizations with 3 -- 4 PCs are feasible since for high-energy lepton fluxes the very fine details of the cosmic ray spectrum are not crucial.\label{fig:GSF_PC}}
\end{figure}
\begin{figure}
    \includegraphics[width=\columnwidth]{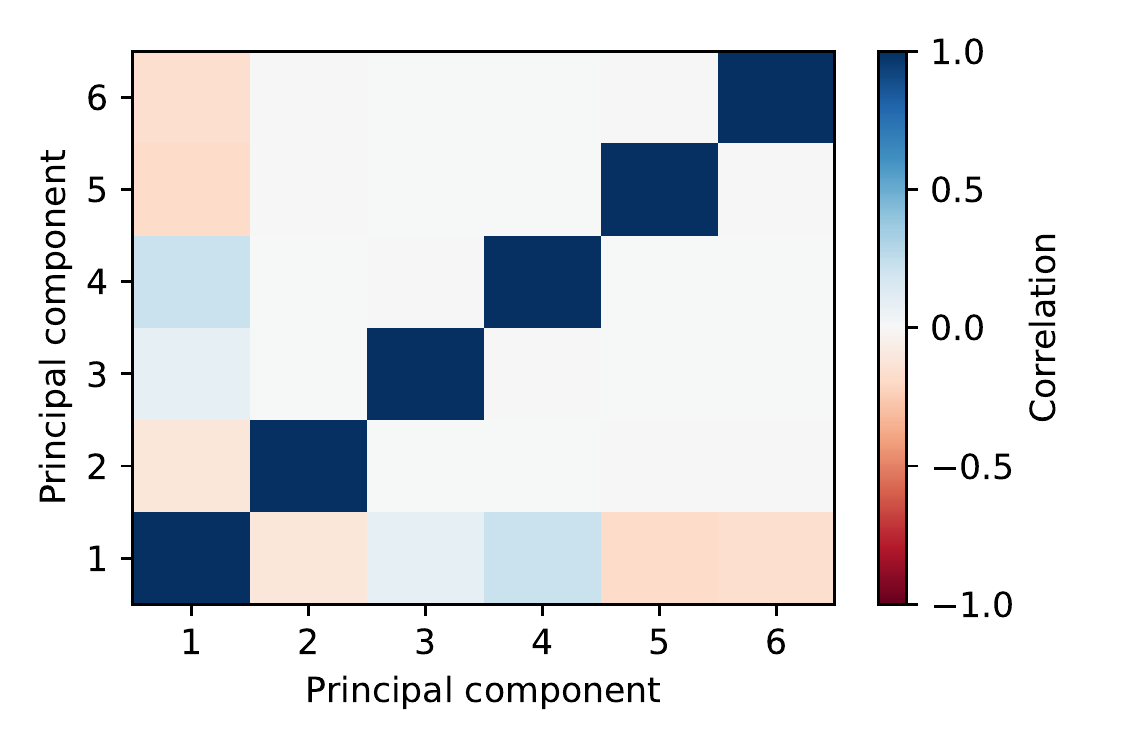}  
    \caption{The correlation matrix of the PCs. By construction, the correlations the PCs should be almost orthogonal but due to the imperfection of the {\sc SVD}, some small correlations remain for the first component.\label{fig:GSF_PC_correlation_matrix}}
\end{figure}
\subsubsection{Dimensionality reduction of GSF}

The parameters of GSF are reduced to a manageable number by means of principal component analysis (PCA). PCA is a statistical method for reducing the complexity of a dataset. The goal is to transform the data into a new coordinate system, where the covariance matrix of the new parameters, referred to as the principal components (PCs), is diagonal. The magnitude of each PC represents how much of the total variance of the data is explained by each component. This technique is widely used and described in more detail in references such as \cite{jolliffe2002principal}.

To address the issue posed by the large numerical range of the covariance matrix elements in the spline coefficients of GSF, which are fitted to the steeply falling cosmic ray flux in linear units, a transformation to logarithmic space was performed. This was achieved by first randomly sampling 10,000 realizations (see Fig.~\ref{fig:GSF_samples}) of the cosmic ray proton and neutron flux, which are required by the semi-superposition approximation in \mceq{}. The proton and neutron components were then concatenated into a single dimensional vector, and the logarithm of each realization was taken. A Singular Value Decomposition (SVD) transformation was then applied to the resulting ensemble, and the dimensions of the factorized matrices were truncated to six components, chosen for this study.

The choice of six components for the GSF-PCA6 model is motivated by the fact that it explains almost 90\% of the variance in the relevant energy range. This can be seen from the explained variance plot shown in Fig.~\ref{fig:GSF_PC}. The majority of the variance is accumulated by features around the knee and higher energies. It is worth noting that the number of components required to explain the variance would be different if the truncation of the energy range was changed to lower or higher energies.

The correlation matrix of the reduced model is shown in Fig.~\ref{fig:GSF_PC_correlation_matrix}. It can be seen that the first component is slightly correlated with the others, whereas the higher components are uncorrelated, as desired by the construction method. The remaining correlations can be attributed to the numerical imperfections in the SVD used to obtain the GSF-PCA6 model.

\begin{figure}
    \includegraphics[width=\columnwidth]{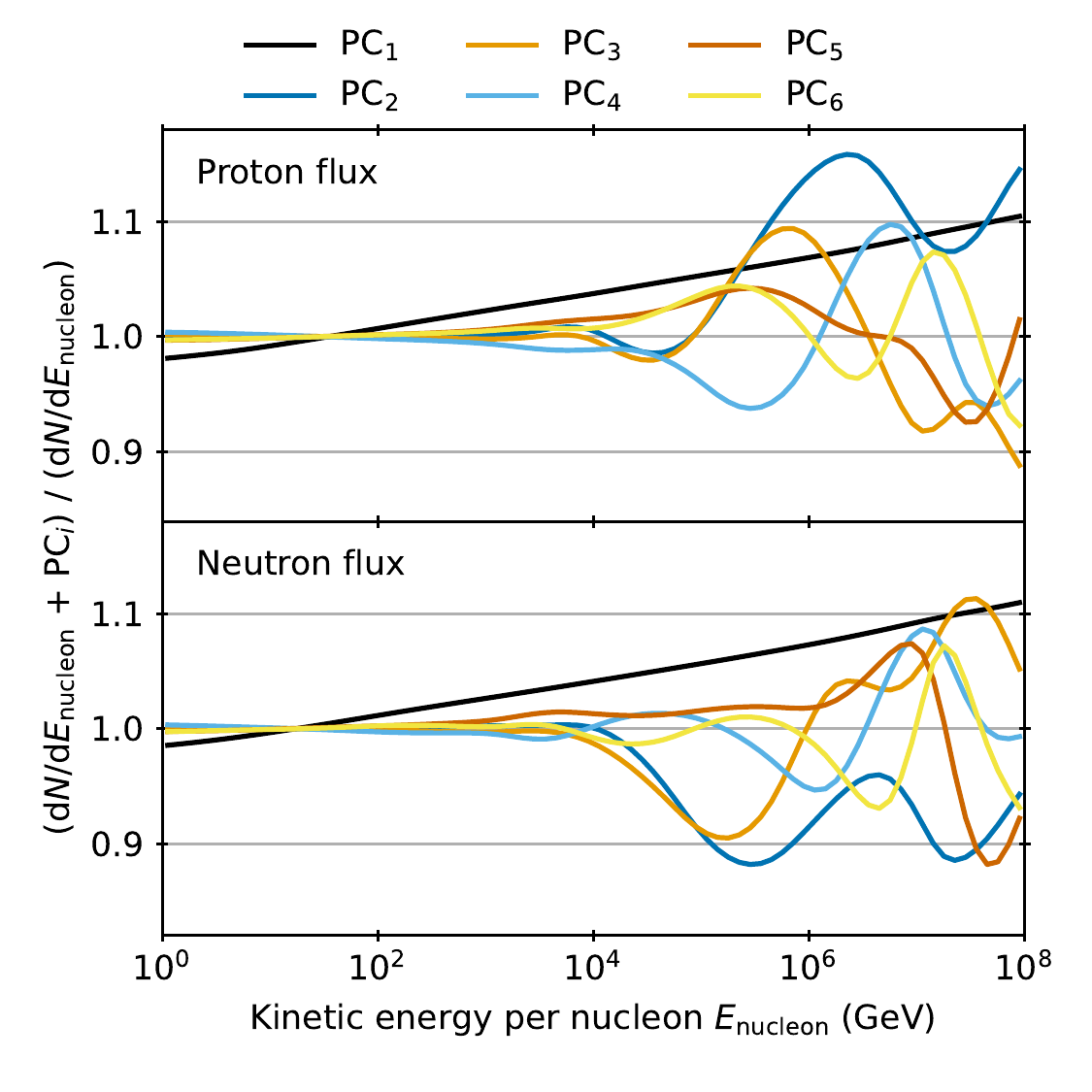}  
    \caption{Spectral features of the six PCs of GSF-PCA6. The panels show the ratio of the total nucleon fluxes, where each component is modified by $+1\sigma$. The leading component, ${\rm PC}_1$, is a uniform spectral index correction for proton and neutron fluxes. Other PCs show (anti-)correlations between proton and neutron, that physically originate from fitting the all-particle and elemental spectra within the GSF approach. At energies of the knee, the other PCs significantly deviate from a power-law spectrum, indicating that a further simplification of the cosmic ray flux errors requires careful assessment.\label{fig:GSF_PC_impact}}
\end{figure}
\begin{figure}
    \includegraphics[width=\columnwidth]{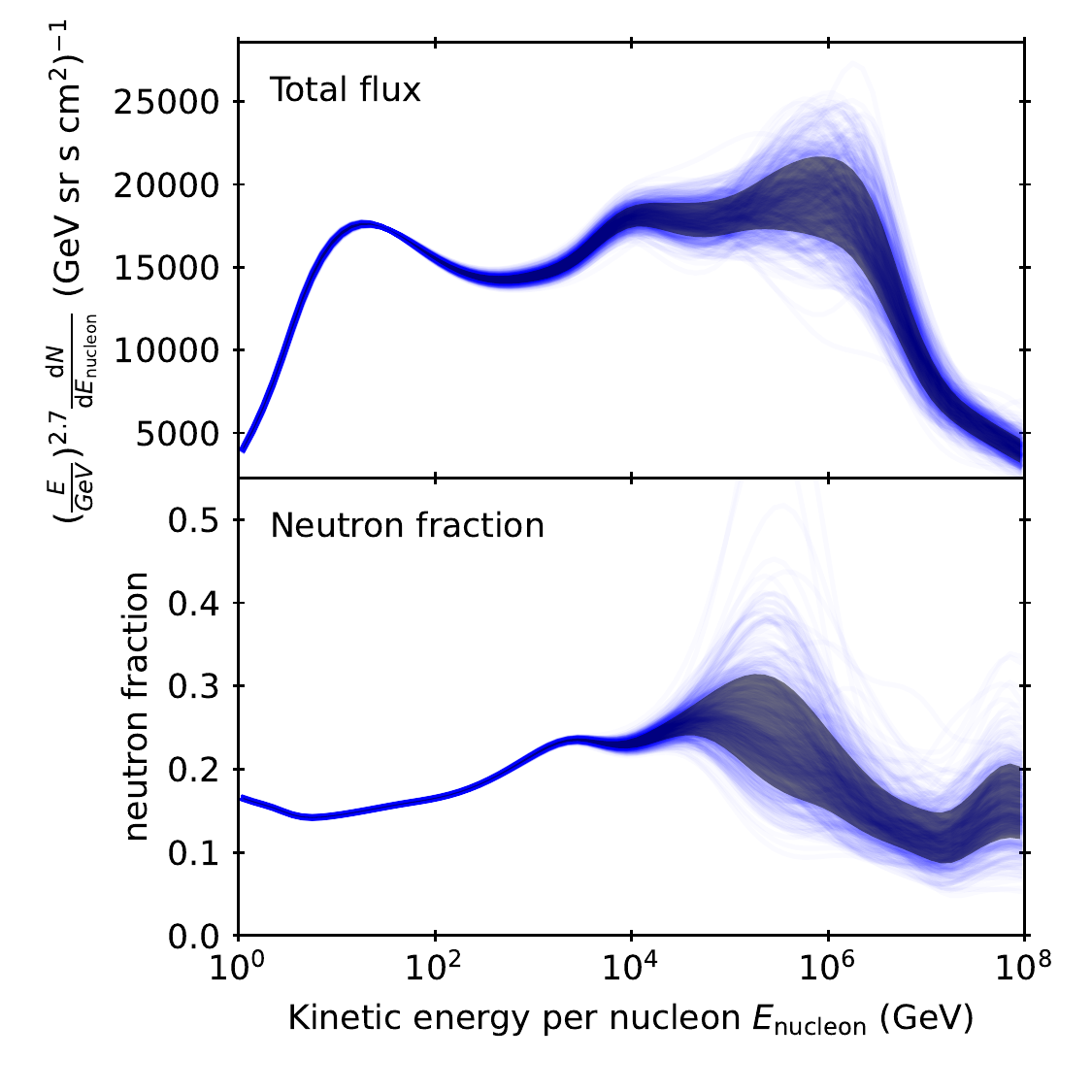}  
    \caption{Realizations of the total nucleon flux and the neutron fraction from GSF-PCA6 with the $1\sigma$ error band superimposed as gray shaded area.\label{fig:reduced_GSF_total_nfraction}}
\end{figure}

\subsubsection{GSF-PCA6}

The six PCs of GSF-PCA6 can be used to alter the cosmic ray nucleon flux within its uncertainties. 
Figure~\ref{fig:GSF_PC_impact} illustrates the impact of varying the PCs within their $1\sigma$ range. Is is remarkable that the method extracts a simple spectral index correction with a pivot point around 80 GeV as the dominant component. The higher PCs are responsible for generating the various allowed spectral shapes in the less certain ranges of GSF around the energies of the knee.

The correlations between the proton and neutron components are conserved from the original model and can be seen by following the directions the components, which tend to correct in opposite directions (e.g., PC$_2$ is positive for protons but negative for neutrons in the upper and lower panel of Fig.~\ref{fig:GSF_PC_impact}, respectively). Some random realizations of GSF-PCA6 are shown in Fig.~\ref{fig:reduced_GSF_total_nfraction} together with the $1\sigma$ error band obtained from regular error propagation. We observe that the uncertainty for the nucleon flux begins to rise above TeV energies, whereas the neutron fraction ($n/(p + n)$) is very well constrained up to $\sim10$ TeV. Above this energy, the mass composition uncertainties of the indirect cosmic ray observations result in large uncertainties of $>30$\%. New, more precise measurements of the cosmic ray composition at PeV energies will be crucial to further reduce the remaining uncertainties.

\subsection{Implementation of calibration parameters}
\label{sec:impl_cal_params}
\mceq{} implements functions to easily change the cosmic ray flux and the hadronic production cross section matrices required to compute lepton fluxes. The latter can be entirely replaced by new cross section matrices or modified starting from a baseline model. Here, we use a scheme that was originally implemented for the propagation of \textit{Bartol} uncertainties \cite{Barr:2006it} in \cite{Fedynitch:2018vfe} and is described in Appendix A of Ref.~\cite{Fedynitch:2022vty}. Its main purpose is to compute the gradients of the lepton fluxes with respect to parameter variations of the cosmic ray flux or the hadronic models without the need of re-running \mceq{} at every parameter change. Below, we explain the two schemes, which can be used for error propagation or for parametrizing the corrections to the lepton fluxes given a set of tuning or nuisance parameters.

\subsubsection{The ``rigorous'' scheme}

This scheme is used to propagate the uncertainties from the accelerator data, which forms the foundation of the \ddm{} model. Also, the same scheme is applied for the propagation of uncertainties from GSF-PCA6.

We compute a muon flux ($\Phi_\mu(E_\mu)$) gradient via first-order finite differences with respect to a variation $\delta$ on single parameter $\mathcal{B}_i$ as
\begin{equation}
    \frac{\partial \Phi_\mu(E_\mu)}{\partial \mathcal{B}_i} = \frac{\Phi_\mu(E_\mu,(1 + \delta)\mathcal{B}_i) - \Phi_\mu(E_\mu,(1 - \delta)\mathcal{B}_i)}{2 \delta}.
\end{equation}
These gradients are interpolated in muon energy (or momentum) and tabulated together with the unperturbed flux for each experimental condition (zenith, atmosphere, season, altitude). The parameters $\mathcal{B}_i$ in the case of \ddm{} are the 75 spline coefficients for its ten differential cross sections. In the case of GSF-PCA6, these are the six PCs. The variations $\delta$ are chosen to be the $1 \sigma$ errors of each spline knot. The flux (or the charge ratio) for a given location under different corrections $\mathcal{B}_i$ can be calculated from
\begin{equation}
    \Phi_\mu(E_\mu, \mathcal{B}_a, \mathcal{B}_b, \dots) = \Phi_\mu(E_\mu) + \sum_i \mathcal{B}_i  \frac{\partial \Phi_\mu(E_\mu)}{\partial \mathcal{B}_i}.
\end{equation}
Since the coupled cascade equations (see for instance \cite{Gaisser:2016uoy}) are linear (with non-constant coefficients), this approach is sufficiently exact and usually does not require higher order terms. 

The flux gradients with respect to the spline knots $\mathcal{B}_i$ are arranged into a Jacobian matrix $\boldsymbol{J}_{\mathcal{BD}}$, which is used together with the covariance matrix $\boldsymbol{\Sigma}_\mathcal{B}$ (obtained in \ddm{} from the fit to accelerator data) to propagate the hadronic cross section flux uncertainty $\delta\Phi_\mu(E_\mu)_\mathcal{D}$ for one spline $\mathcal{D}$ (corresponding to a single differential cross section):
\begin{equation}
    \delta\Phi_\mu(E_\mu)_\mathcal{D} = \boldsymbol{J}_{\mathcal{BD}}^T \boldsymbol{\Sigma}_\mathcal{B} \boldsymbol{J}_{\mathcal{BD}}.
    \label{eq:uncertainty_per_spline}
\end{equation}
The process is repeated for each experimental location and zenith angle bin. However, the published fixed-target data only allows for the error propagation to be performed independently for each of the different splines $\mathcal{D}$, such as $\pi^+$ or $\pi^-$ yields, without taking into account potential correlations between the measurements, such as the pion charge ratio or kaon-to-pion ratio. This limitation is due to the unavailability of single-differential data binned in the required $\xl$ variable, and not a limitation of the method, which could incorporate a larger covariance matrix that includes correlations among different particle species. 

In this work, the error propagation reduces the number of free parameters from the 75 spline knots to the number of independent cross sections, ten by default, denoted as $\mathcal{D}_\lepip$, $\mathcal{D}_\hepip$, etc. The indices, such as 31G and 158G, in the previous example are a combination of the beam energy and the first letter of the energy unit (GeV in this case).

Once the resulting database of fluxes and gradients for each experimental site is pre-calculated, the evaluations of arbitrary combinations of the $\mathcal{D}$ parameters are very fast and can be directly used by minimizers. A new flux model can be readily computed from
\begin{multline}
        \Phi_\mu(E_\mu, \mathcal{D}_\lepip, \dots, \mathcal{D}_\GSFVI) \\
        = \Phi_\mu(E_\mu) + \sum_i \mathcal{D}_i \, \delta\Phi_\mu(E_\mu)_\mathcal{D}.
\label{eq:new_flux}
\end{multline}
The values $\mathcal{D}_i$ for $i = \lepip, \lepim, \dots, \heKp, \dots$ and the covariance matrix $\boldsymbol{\Sigma}_\mathcal{D}$ are obtained by fitting the data. Arranging the gradients $\delta\Phi_\mu(E_\mu)_\mathcal{D}$ defined in Eq.~\ref{eq:uncertainty_per_spline} into a new Jacobian matrix $\boldsymbol{J}_{\mathcal{D}}$ gives the total error after an additional error propagation step
\begin{equation}
    \delta\Phi_\mu(E_\mu) = \boldsymbol{J}_{\mathcal{D}}^T \boldsymbol{\Sigma}_\mathcal{D} \boldsymbol{J}_{\mathcal{D}}.
    \label{eq:uncertainty_after_fit}
\end{equation}
The calibrated neutrino fluxes are calculated in the same way using the central fluxes and gradients for a specific neutrino flavor and the same $\mathcal{D}$s and $\boldsymbol{\Sigma}_\mathcal{D}$ obtained from fitting the muon data.

The method referred to as rigorous distinguishes itself by its comprehensive approach to calculating gradients that reflect the uncertainty in the DDM. This uncertainty is derived from the $1\sigma$ errors of each spline knot, with $\mathcal{D}=\pm1$ representing the error in the accelerator data. While this method offers a clear understanding of the error, it's worth noting that its definition of positive error results in positive gradients. This may not always provide an accurate representation, particularly in cases like charge ratios where an increase in negative hadrons results in a negative correction. The calculation of charge or flavor ratios from the quotients of corrected fluxes using Eq.~\ref{eq:new_flux} would not be technically correct under the Taylor expansion. As a result, the `rigorous' method may be effective in analyzing fluxes, but not ratios. Additionally, the definition of the fit parameters $\mathcal{D}$ as a scale of the flux gradients, instead of at the cross-section level before the cascade equations are solved, makes it difficult to implement a ``tuned" model in \mceq{}. This requires the use of a tabulated representation rather than a direct representation of the best fit fluxes.

\subsubsection{The ``practical'' scheme}
An approximate, computationally simpler scheme derives the gradients with respect to the $\mathcal{P}$ parameters from a simple scale perturbation of secondary particle yields ($\rmd N^\mathcal{P}/\rmd \xl)$:
\begin{equation}
\label{eq:simple_gradients}
	\delta\Phi_\mu(E_\mu)_\mathcal{P} = \frac{\Phi_\mu(E_\mu,(1 + \delta)\frac{\rmd N^\mathcal{P}}{\rmd \xl}) - \Phi_\mu(E_\mu,(1 - \delta) \frac{\rmd N^\mathcal{P}}{\rmd \xl})}{2 \delta}.
\end{equation}
The disadvantage of this approach is that it fails to explicitly account for the decrease in cross-section error from the accelerator data as $\xl$ becomes smaller. To address this, an equivalent $1\sigma$ scale for the $\mathcal{P}$-parameters can be established by equalizing the errors of spectrum-weighted moments (Z-factors) between the schemes:
\begin{equation}
\label{eq:zfactor}
	\delta Z_{{\rm p}\mathcal{D}_i} \equiv \mathcal{P}_i \int_0^1 \rmd \xl \, x^{1.7} \frac{\rmd N^\mathcal{P}}{\rmd \xl}.
\end{equation}
The Z-factor error $\delta Z_{{\rm p}\mathcal{D}_i}$ is calculated from the \ddm{} splines by standard error propagation. This approach ensures that $\mathcal{P}_i = \pm 1$ approximately corresponds to $\mathcal{D} = \pm1$, which is the propagated $1\sigma$ error of the accelerator data.

\subsubsection{Choice of high-energy extrapolation parameters}

The hadronic yields in the DDM are assumed to follow Feynman scaling. However, to increase flexibility at projectile energies $E_{\text p} > 158$ GeV, additional parameters are introduced as part of this work. This is done by ``cloning" the 158 GeV yields ($\rmd N/\rmd \xl$) at higher projectile energies and linearly interpolating between these points in $\log{E_{\text p}}$. These additional degrees of freedom allow us to test potential deviations from Feynman scaling and quantify the extrapolation errors of the muon and neutrino fluxes above TeV energies. 

Altering the choice of the very-high energy parameters (denoted by $^\bigstar$) requires re-generating the gradient library, as the energy interpolation changes the shape of the flux gradients. This provides a valuable tool for determining the extent to which the muon data constrains the muon and neutrino fluxes, rather than relying solely on rigid model assumptions.

\begin{figure*}
    \centering
    \includegraphics[width=\textwidth]{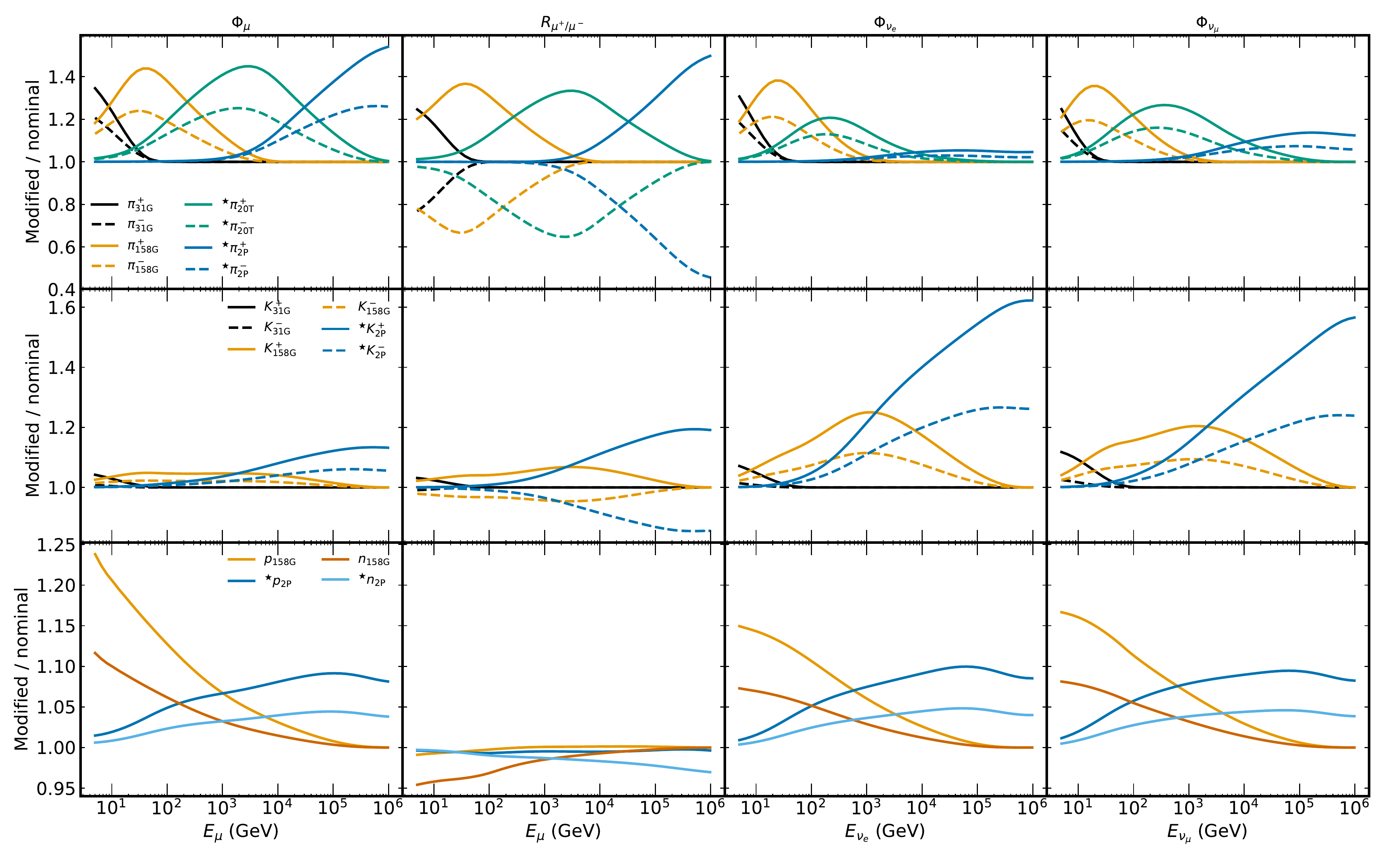}
    \caption{
    This figure shows the effect of the gradients divided by the flux, $\delta\Phi_\mu(E_\mu)_\mathcal{P}/\Phi_\mu(E_\mu)_\mathcal{P}$ (see Eq.~\ref{eq:simple_gradients}). The modifications made by positively charged hadrons are represented by solid lines, while those made by negatively charged hadrons are indicated by dashed lines.
    }
    \label{fig:impact_hadrons}
\end{figure*}

\begin{figure*}
\centering
\includegraphics[width=\textwidth]{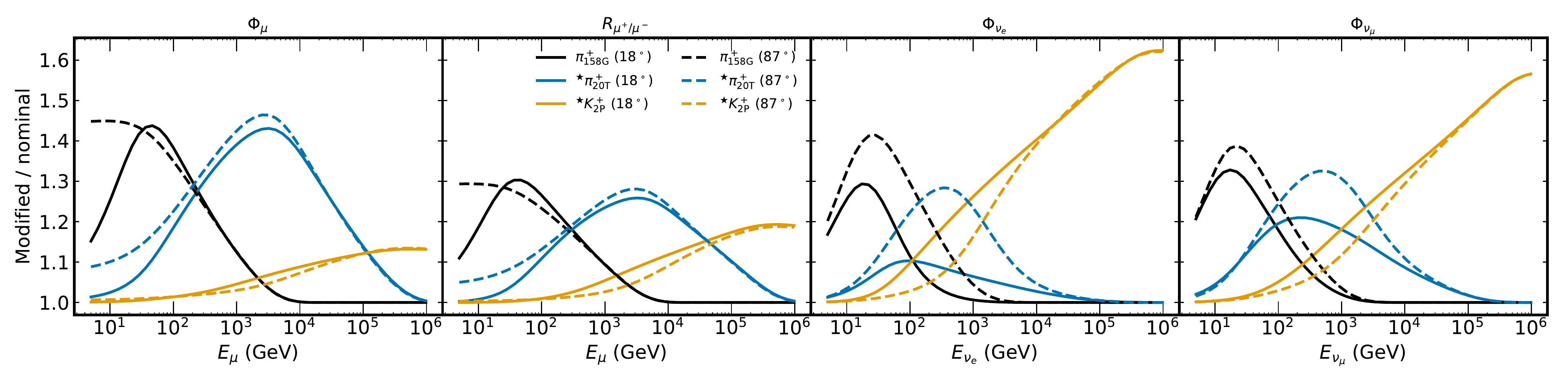}
\caption{This figure shows the effect of the gradients divided by the flux, $\delta\Phi_\mu(E_\mu)_\mathcal{P}/\Phi_\mu(E_\mu)_\mathcal{P}$ for two different zenith angles. It emphasizes the crucial role of accounting for the zenith angle dependence accurately when comparing to experimental observations.}
\label{fig:impact_zenith}
\end{figure*}

We assessed the effect of adding parameters for all particle yields at 2 TeV, 20 TeV, 200 TeV, and 2 PeV on the muon flux and charge ratio predictions. Our results revealed that a gap of at least two decades in energy was necessary to prevent strong correlations. The majority of sensitivity was found to be in the pion yields, while variations in other particles had a limited impact on the muon flux. As a result, we retained two additional calibration points for pions at 20~TeV and 2~PeV, and one point at 2~PeV for the rest of the hadronic yields.

The results of this scheme are depicted in Figures~\ref{fig:impact_hadrons} and \ref{fig:impact_zenith}, which show the gradients computed using Eq.~\ref{eq:simple_gradients}. These figures reveal the impact that the gradients have on the flux, as well as the muon charge ratio. As shown, the largest impact on muon fluxes is from $\pi^+$ across all energy ranges. This impact is more pronounced than that of $\pi^-$ due to the greater abundance of $\pi^+$ in the atmosphere. Our fit is expected to have reduced sensitivity to kaon production in the atmosphere, potentially causing significant errors in high-energy neutrino predictions as these are primarily driven by kaon decays at TeV energies. Changes to the cosmic ray parameters impacting the spectrum have similar effects on muon and neutrino fluxes as shown in Fig.~\ref{fig:GSF_PC_impact}, but minimal impact on the muon charge ratio.

\section{Selection and treatment of muon datasets}
\label{sec:data}

We surveyed the literature to collect all published measurements of muon fluxes and charge ratios by spectrometers as a function of energy or momentum, and studied the subset with muon momentum higher than 5\,GeV at the detector to avoid complications introduced by geomagnetic effects, not yet included in \mceq{}. Our survey included the comprehensive list of historical measurements presented in~\cite{Fiorentini:2001wa} as well as the modern measurements reported in~\cite{ParticleDataGroup:2022pth}.

While there are numerous data sets of atmospheric muon fluxes, the level of detail given on their systematic uncertainties is often insufficient to model their measurements with percent-precision. This applies both to the performance of the detector as well as the assumptions made while analyzing the data. This is further complicated by the common practice done by most experiments to transform their data into a vertical-equivalent measurement at the surface, applying corrections that make the assessment of their systematic uncertainties challenging. For these reasons we restrict ourselves to use only data sets with a detailed description of either the corrections used during analysis or the resulting uncertainties introduced by them. 

A key aspect of our fit procedure is that, whenever possible, we account for the systematic uncertainties of each experiment by introducing correction functions that modify the reported measurements. We were able to do this for BESS-TeV, L3+c and MINOS. We note that these corrections are necessary to bring the vertical fluxes of BESS-TeV and L3+c into agreement.

When selecting the data sets we also introduce a requirement for the data to be consistent, within errors, with the expected properties of the muon flux (see Sec.~\ref{sec:flux_test}). We found a total of seven experiments that fulfilled all the requirements: BESS-TeV, CMS, DEIS, L3+c, MINOS, MUTRON and OPERA. They are summarized in Table~\ref{table:experiments} and discussed in further detail below.

\begin{table*}[!t]
\centering

\begin{tabularx}\linewidth{p{0.15\linewidth}p{0.13\linewidth}p{0.13\linewidth}p{0.05\linewidth}p{0.10\linewidth}p{0.16\linewidth}p{0.09\linewidth}l}
\hline
Experiment & Energy (GeV) & Measurements          & Unit & Systematics & Location               & Altitude        & Zenith range                           \\ \hline
BESS-TeV \cite{Haino:2004nq}   & 0.6-400            & \muflux                  & $p_\mu$   & C  & 36.2$^\circ$N, 140.1$^\circ$W        & 30 m            & 0-25.8$^\circ$                 \\
CMS~\cite{Khachatryan:2010mw}        & 5-1000             & \muratio         & $p_\mu$    & Q & 46.31$^\circ$N, 6.071$^\circ$E       & 420 m           & $p\cos\theta_z$                                         \\
L3+C \cite{Achard:2004ws}       & 20-3000            & \muflux,\muratio & $p_\mu$    & C & 46.25$^\circ$N,  6.02$^\circ$E       & 450 m           & 0-58$^\circ$                                                                     \\
DEIS \cite{Allkofer:1985ey}       & 5-10000            & \muratio    & $p_\mu$    & Q & 32.11$^\circ$N, 34.80$^\circ$E &  5 m &  78.1-90$^\circ$ \\
MUTRON \cite{Matsuno:1984kq}       & 80-10000            & \muratio    & $p_\mu$    & Q & 35.67$^\circ$N, 139.70$^\circ$E &  5 m &  87-90$^\circ$ \\
MINOS \cite{Adamson:2007ww}      & 1000-7000          & \muratio          & $E_\mu$ & C & 47.82$^\circ$N, 92.24$^\circ$W & 5 m  & unfolded                           \\
OPERA \cite{Agafonova:2014mzx}      & 891-7079           & \muratio          & $E_\mu$ & Q & 42.42$^\circ$N, 13.51$^\circ$E   & 5 m & $E\cos\theta^*$                                 \\ \hline
\end{tabularx}
\caption{List of measurements used in the flux calibration. Most data are either taken for vertical incidence angles or corrected to vertical through model-dependent unfolding. The systematics column indicates whether errors were added in quadrature (Q) or as correction functions (C). For the DEIS data set only the charge ratio was used in our final fit. See text and Appendix~\ref{appendix:deis} for details. The measurements from MINOS and OPERA are reported as surface-equivalent fluxes by the collaborations, and are therefore compared with a calculation at an altitude of 5~m. }
\label{table:experiments}
\end{table*}

\subsection{BESS-TeV}

The BESS-TeV spectrometer was a cylindrical detector using a thin superconductor solenoid~\cite{BESS:2000gvp, Haino:2004gx}. Thanks to this geometry, BESS-TeV had an almost constant geometrical acceptance and uniform performance for various incident angles and positions. It used a 1T magnetic field to deflect particles as they crossed it, measuring them with five different drift chambers.

Two observations of muon fluxes and charge ratios from BESS-TeV have been reported~\cite{Haino:2004nq}. In one of them, the spectrometer was launched by a balloon in Manitoba, Canada, and took data for 10.6 hours at an altitude of 37~km. The ground observation was done at KEK, in Tsukuba, Japan. The muon flux data reported is for vertically downgoing particles, requiring $\cos\theta_z > 0.98$ for rigidities $R<20$~GV and $\cos\theta_z>0.90$ for $R>20$~GV. While BESS-TeV reports that the choice of cutoff angle did not change their observed flux, their data is precise enough that our calculations were sensitive tho this effects. For this reason, our work uses different zenith to analyze BESS-TeV data.

BESS reports systematic uncertainties per data point, as well as two corrections to the spectrum due to the finite resolution and potential misalignment of the spectrometer. In this work we consider both sources of error, adding the point-by-point errors in quadrature to the statistical one, and using the misalignment and resolution functions as corrections to potential biases. These spectral deformations were taken from~\cite{Haino:2004nq}, where the ratio of the systematically modified flux to the original one $\Phi^\mathrm{sys}_\mu(p)/\Phi_\mu(p)$ is given as function of energy. We interpret the modified flux to be $\Phi^\mathrm{sys}_\mu(p) = \Phi_\mu(p) + \frac{\delta \Phi_\mu(p)}{\delta S}$, where $S$ represents a systematic change in either alignment or resolution. The adjusted flux is then given by
\begin{equation}
\Phi'_\mu(p) = \Phi_\mu(p) + S_\mathrm{al} \frac{\delta \Phi_\mu(p)}{\delta S_\mathrm{al}}+ S_\mathrm{res} \frac{\delta \Phi_\mu(p)}{\delta S_\mathrm{res}},
\end{equation}
where \textit{al} and \textit{res} stand for alignment and resolution, respectively, and the $S$ coefficients can be adjusted. The reported global acceptance error of 0.3\% is also added in quadrature to the statistical errors, which are at best 1.5\%.

\subsection{CMS}

The Compact Muon Sollenoid is a detector situated along the beam line of the Large Hadron Collider at CERN, on the border between Switzerland and France, at a depth of 89~m~\cite{CMS:2008xjf}. Its main feature is a superconducting magnet that produces a magnetic field of 3.8~T. Trackers, calorimeters, gas ionization detectors, and resistive plate chambers detect particles as they cross the device.

The data used here is the cosmic-muon charge ratio measured using multiple data sets that rely on different parts of the detector~\cite{Khachatryan:2010mw}. Systematic uncertainties that affect the data, such as variations in the magnetic field or asymmetries in the detector acceptance, have an estimated impact smaller than the statistical uncertainties of the measurement and are already included in the total error reported. The data are therefore used without any modifications.

\subsection{DEIS}
The DEIS apparatus was a muon spectrometer in operation at the University of Tel-Aviv, in Israel, using iron magnets, scintillation counters and wire spark chambers oriented in a way to detect horizontal muons~\cite{deis_proc}. Using a multi-layer configuration, the detector could trace muon tracks as they passed through it. Solid iron magnets were used for particle bending, reaching a field homogeneity better than 1\%. 

The DEIS data set is one of the two measurements of horizontal muons available in the literature. The other one is from MUTRON and is described below. We therefore explored including the muon flux reported in~\cite{Allkofer:1985ey} as well as the muon charge ratio presented in~\cite{Allkofer:1979wk}. Ultimately, only the muon charge ratio was used in the fit, taking the total errors reported by DEIS without further corrections. A detailed discussion of the impact of the DEIS muon flux data set in our study, including the reasons for not including it in the final result, can be found in Appendix~\ref{appendix:deis}.

\subsection{L3+C}

The L3+C experiment~\cite{Adriani:2002uu} was installed at the Large Electron Positron Collider at CERN, and consisted of a muon spectrometer below ground and an array of scintillators about 60~m above the axis of the spectrometer. Measurements of cosmic muons were performed using the L3 setup, which detected the arrival of muons using a scintillator array first, followed by concentric drift chambers subject to a magnetic field of 0.5~T. The L3+C collaboration reports measurements of muon fluxes and charge ratio.

The data are subject to three classes of systematic uncertainties that modify the momentum scale $\delta_p$, the flux normalization as function of direction and the detector response matrix~\cite{Achard:2004ws}. The zenith-dependent normalizations arise due to the limited precision of the calibration measurements and the accuracy the detector model in simulation. The uncertainty in the response matrix comes from the limited statistics in the simulation and affects the momentum resolution. The total momentum uncertainty, on the other hand, is a sum of four contributions: the magnetic field, the alignment of the muon chambers, the energy losses in the material, and the modeling of the overburden. The functions that describe the effect of each source of uncertainty on the particle flux and momentum were taken from~\cite{Unger:2004cu}. 

The momentum uncertainty $\delta_p$ results in an uncertainty on the flux, described in~\cite{Dembinski:2017zsh}, as
\begin{equation}
\Phi'_\mu(\theta, p') =  \Phi_\mu\left(\theta, \frac{p}{1+\delta_p}\right)\frac{1}{1+\delta_p}.
\label{eq:l3c_punc}
\end{equation}
To obtain the fully corrected flux the contributions from changing the normalization and momentum resolution are added to the flux of Eq.~\ref{eq:l3c_punc} letting their strength $S$ vary. The modified flux is then given by
\begin{equation}
 \Phi^s_\mu(\theta,p')=\Phi'_\mu (\theta,p') + S_\mathrm{n} \frac{\delta \Phi_\mu(\theta, p')}{\delta S_\mathrm{n}} + S_{p\mathrm{r}} \frac{\delta \Phi_\mu(\theta,p')}{\delta S_{p\mathrm{r}}}.
 \end{equation}
Here \textit{n} and \textit{pr} stand for normalization and momentum resolution, respectively.

\begin{figure}
\centering
\includegraphics[width=0.9\columnwidth]{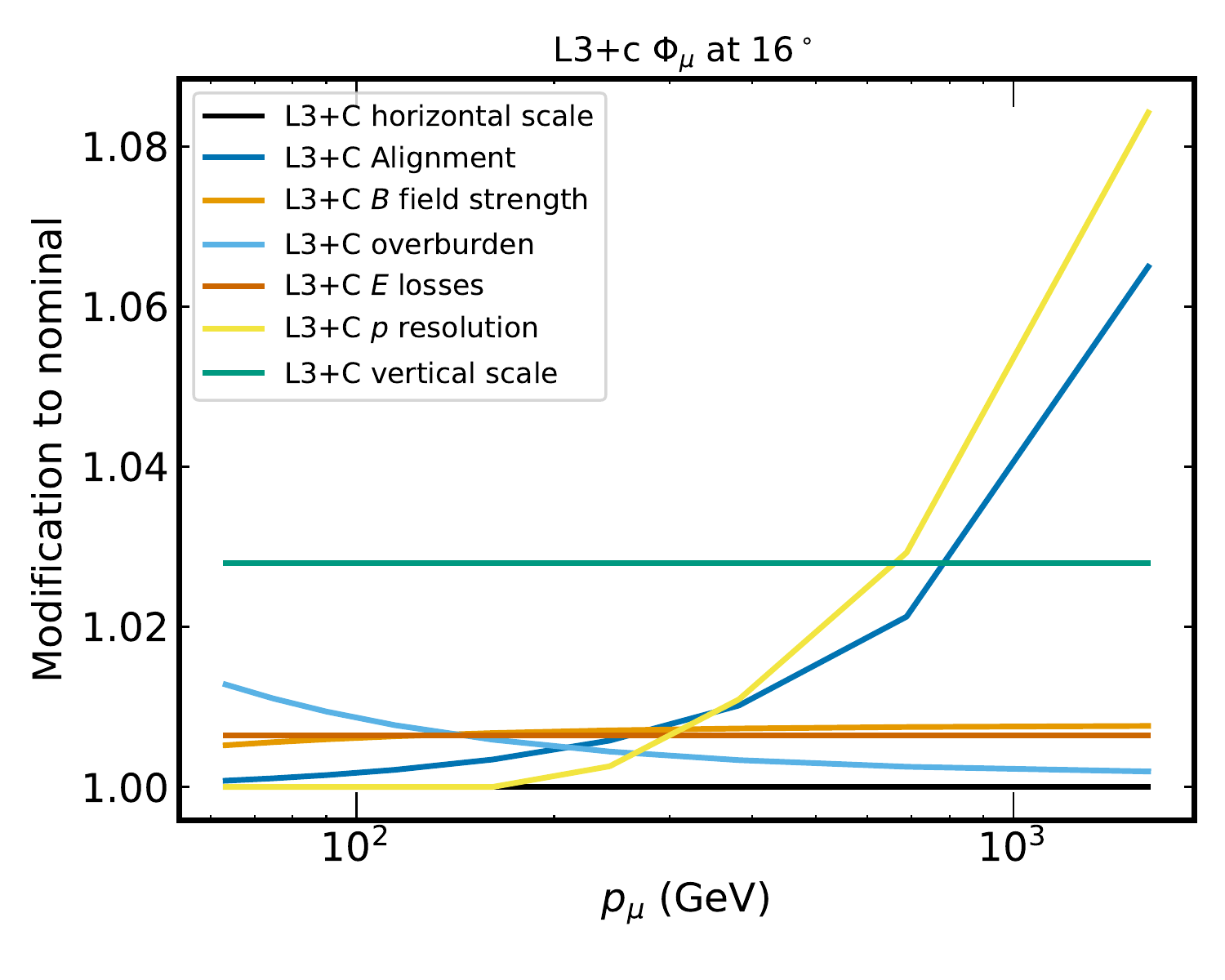}
\includegraphics[width=0.9\columnwidth]{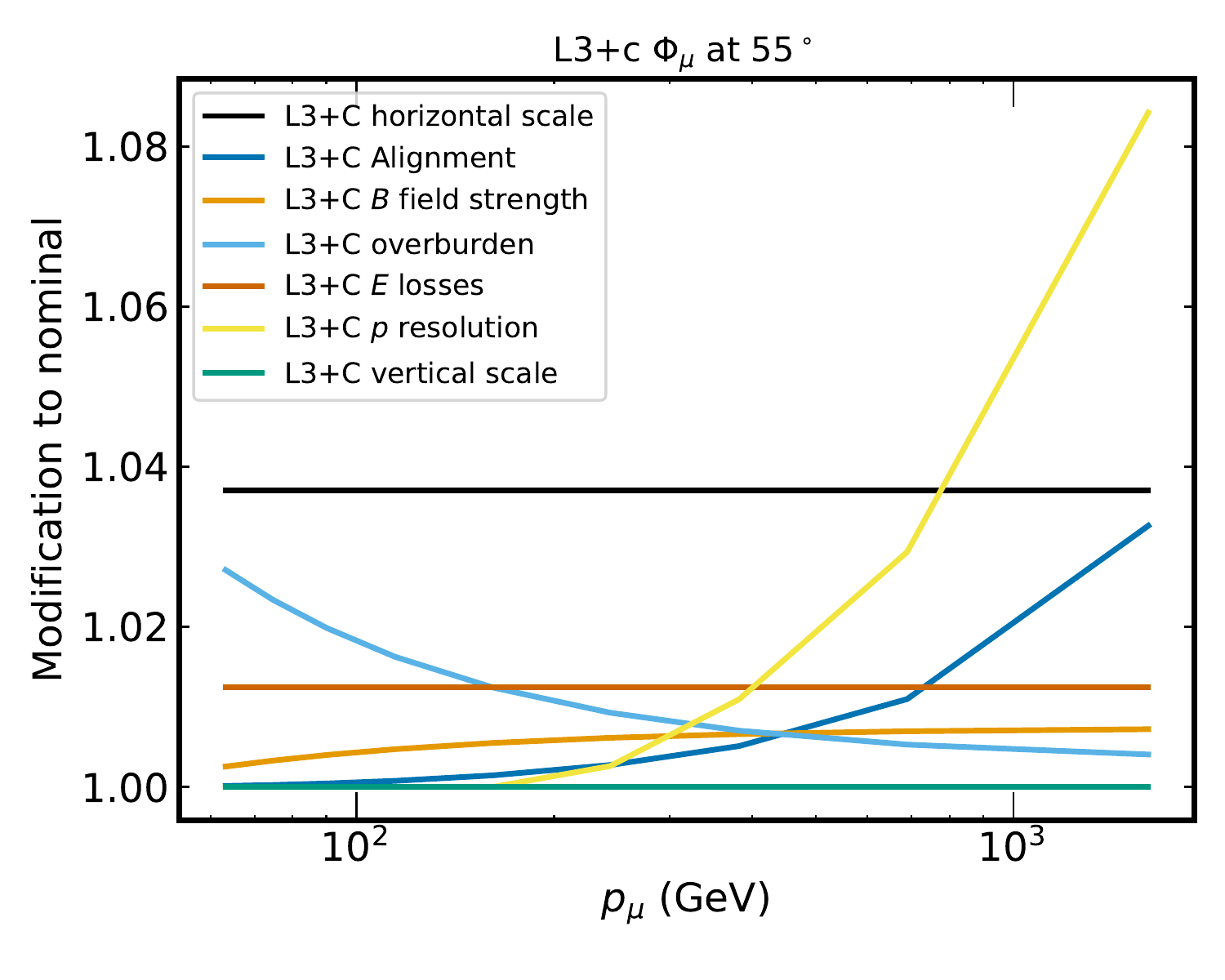}
\caption{Ratio of original and perturbed flux after modifying each source of uncertainty by $1\sigma$ for the measurement at 16\degrees{} (top) and 55\degrees{} (bottom) by L3+C. The figures only show the response that produces higher fluxes, but the parameters are symmetric around one and can reduce the flux in a similar fashion.}
\label{fig:l3c_syst}
\end{figure}

The corrections are energy and zenith-angle dependent and, in the case of the alignment of the muon chambers, charge dependent. Figure~\ref{fig:l3c_syst} shows the impact of each correction for two angles. Even though the strength varies, in both cases the overburden is the most relevant at lower energies, while the momentum resolution is the leading source of error at the highest energies.

\subsection{MINOS}

The MINOS far detector was a steel-scintillator calorimeter situated at about 2070 meter water equivalent (m.w.e.) in the Soudan Underground Laboratory in Minnesota, USA~\cite{MINOS:2008hdf}. The detector, originally designed to record neutrino interactions, consists of steel plates interleaved with plastic scintillator strips. The steel was magnetized into a toroidal field configuration, with an average magnetic field strength of 1.3~T. 

For this study we used the cosmic muon ratio data sample collected by MINOS between 2003 and 2006~\cite{Adamson:2007ww}. The main systematic uncertainty for this result is the energy scale of the measurement, which is reported at the surface and thus was corrected for overburden. This uncertainty, estimated to be 20\%, is introduced as a correction factor during the fit that changes the energy associated with the measurement. As for rest of the experiments, the muon charge ratio data is rather robust against detection uncertainties. The combined effect of other uncertainties, besides the energy scale, was reported to be  smaller than the statistical errors of each data point, and are therefore not considered in our study.

\subsection{MUTRON}
 
MUTRON was a solid iron magnetic spectrometer equipped with multiwire proportional chambers and wire spark chambers, that was set up near Tokyo, Japan~\cite{HIGASHI1978387}. The detector was set up horizontally, with a zenith angle range between 86$^\circ$-90$^\circ$, an angular region that overlaps with that of DEIS.

The experiment measured the absolute flux of muons as well as the muon charge ratio. The description of the uncertainties affecting their measurements, however, is not sufficient to include their flux measurements in our analysis. The muon charge ratio, on the other hand, benefits from cancellation of most systematic uncertainties, in particular for detectors at the surface. For these reasons, we only use the muon charge ratio data in~\cite{Matsuno:1984kq} for our work.

\subsection{OPERA}

The OPERA detector was a hybrid electronic/emulsion experiment situated in LNGS, Italy, at an average depth of 3800 m.w.e.~\cite{KOMATSU2003124}. The detector was designed to detect appearance of $\nu_\tau$ from a $\nu_\mu$ beam, but also took atmospheric muon charge ratio data at TeV-equivalent surface energies~\cite{Agafonova:2014mzx}. The data is transformed to surface-equivalent muon charge ratios by the collaboration, and is thus compared with predictions at 5~m altitude.

The main sources of systematic uncertainties in the measurement of muons are due to alignment and charge misidentification. The errors due to alignment cancel out when reporting the charge ratio, while the charge misidentification results in a modification of the ratio of 0.7\%. We add this uncertainty to the statistical error in quadrature.

\subsection{Flux compatibility test and rejected data sets}
\label{sec:flux_test}

Two different tests were performed on each data set to decide whether (a) their reported errors are normally distributed around some reasonable flux or charge ratio and (b) they are compatible with other data sets. Both of these tests rely on the observation that features in the muon flux or the charge ratio develop over one or more decades of energy, and thus $E^3$\muflux and \muratio are expected to be smooth as a function of $\log_{10}(E/\mathrm{GeV})$ (or momentum).

\begin{figure}[!b]
\centering
\includegraphics[width=1\columnwidth]{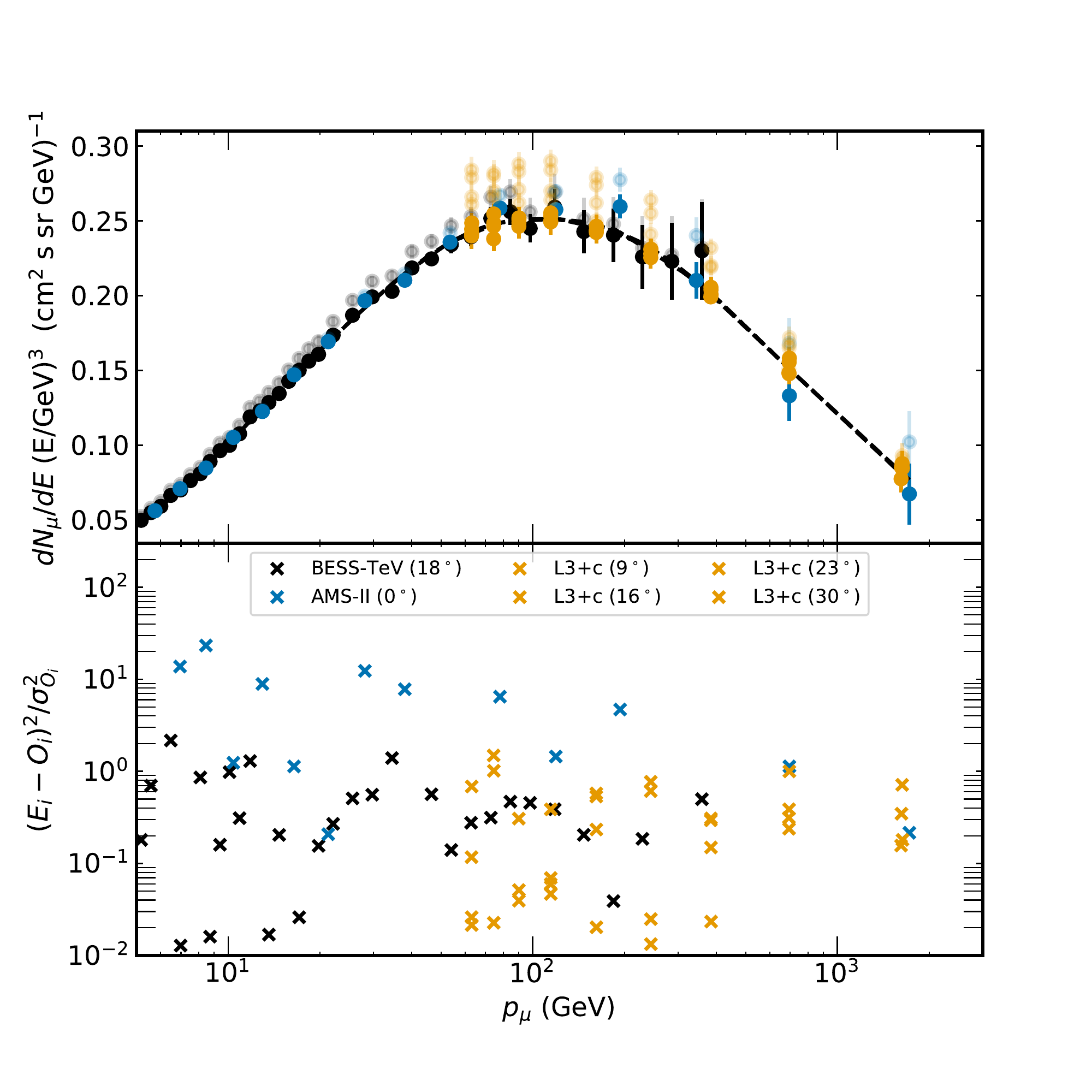}
\caption{Compatibility between vertical fluxes. The top panel shows the fit fluxes in solid colors, after they have been multiplied times $\cos\theta$ to correct for their different inclination. The dashed line that crosses them is the smooth average flux. The bottom panel shows the $\chi^2$ contribution of each data point with respect to the smooth average flux. The AMS-02 measurement has significant deviations below 50~GeV. The variability between adjacent points can be seen in their different $\chi^2$ values at low energies.}
\label{fig:compatibility}
\end{figure}

Since all considered data sets contain multiple data points per decade in energy, we can use these measurements to build a cubic spline with evenly distributed knots to follow the trend of the data and produce a \textit{smooth average flux} or \textit{smooth average charge ratio}. These smooth expectations were calculated from single data sets to test the coverage of their reported errors, as well as from a combination of data sets for vertical directions, correcting by their arrival angle $\cos\theta$, to test the compatibility among them. A $\chi^2$ was computed for each test including systematic uncertainties, varying the systematic correction functions when available, which was used to estimate $p$-values for the data to come from the smooth flux or charge ratio functions. Values smaller than 5\% were studied in detail to decide if data sets were to be included in our study.

The combined fit of vertical fluxes from BESS-TeV and L3+c showed good agreement between the data sets, and was used to test the potential addition of AMS-02 data~\cite{Duranti:2012css}. The data, although not published in a peer-reviewed journal, reports significantly smaller errors than any other data set available and could provide meaningful constraints to the study. However, adding AMS-02 made the agreement between the sets negligibly small. This is shown in Fig.~\ref{fig:compatibility}, where the smooth flux is compared to near-vertical measurements. The AMS-02 data was therefore not included in the fit.

Besides AMS-02 we also considered including muon flux measurements from the MARS~\cite{Ayre:1975qi} and CAPRICE~\cite{WiZardCAPRICE:2000bby,Kremer:1999sg} experiments. Both measured muon fluxes at vertical directions, with a maximum momentum around 500~GeV/$c$. The uncertainties of MARS are well described but, upon inclusion, we found it to to require very significant pull on its systematics errors to agree with L3+C, and that it would disagree with BESS-TeV regardless of the use of correction functions. When combined with both measurements, the fit using the smooth flux method returned a negligible small $p$-value. The data from CAPRICE94 and CAPRICE97 are reported only with total errors, so no additional treatment for systematic uncertainties was given to it. Their data is not well described by a smooth flux and it introduces tensions among the other data sets, failing the compatibility criterion.

We performed similar studies on the near-horizontal muon flux data sets from MUTRON and DEIS and found that neither of them could be described by a smooth flux. Since these are the only near-horizontal sets, we carried out further tests on the DEIS data. The final result, however, does not include this set as it introduced significant tensions with the rest of the experiments, as discussed in Appendix~\ref{appendix:deis}.

The muon charge ratio data was also tested for compatibility and smoothness. These data are rather featureless as a function of energy and at low energies most of its systematic errors cancel out. As a result, all experiments passed our smoothness and compatibility tests.

\section{Analysis and Results}

\subsection{Fitting procedure}

The calibration of the lepton fluxes is done by fitting the muon fluxes and the \muratio{} produced by \mceq{} to data by varying 34 parameters: six from the GSF-PCA6 model, 18 from the DDM including the extrapolation parameters, and 10 are corrections to the experimental uncertainties of the data in the fit. The test statistic is a modified $\chi^2$ given by
\begin{equation}
    \chi^2 = \sum_i^N \frac{(O_i - T_i)^2}{\sigma_{O_i}^2} + \sum_j^M \frac{(F_j - G_j)^2}{\sigma_{F_j}^2} 
   \label{eq:modchi2}
\end{equation}
where the first sum compares observation $O_i$ at each data point $i$ with expectation $T_i$ from \mceq{} divided by the error in the observation $\sigma_{O_i}$. The second sum accounts for prior knowledge on each $M$ parameter, penalizing deviations $F_j$ from their expectation $G_j$ divided by its estimated error $\sigma_{F_j}$. At each iteration of the fit, all experimental data sets are modified and the expectation is recomputed using Eq.~\ref{eq:new_flux}.

The $M$ parameters that modify the cosmic ray flux and hadronic interaction models impact all data sets simultaneously, although their strength depends on the type of data, energy and observed directions of each experiment. All parameters are varied simultaneously during the minimization of Eq.~\ref{eq:modchi2} by i{\sc Minuit}~\cite{iminuit}. 

Most free parameters have a well defined expectation ($G$) and uncertainty ($\sigma_{F_j}$) with the exception of the extrapolation $\star$-ed parameters, which are not constrained by accelerator data. In those cases, we preserve the shape and central value of the highest energy differential cross section, and increase the error until the Z-factors predicted by four modern different hadronic interaction models fall within its range. The models used for this comparison are \sibyll{}-2.3d, DPMJET-III 19.1 \cite{Roesler:2000he,Fedynitch:2015kcn}, EPOS-LHC \cite{Pierog:2013ria} and QGSJET-II-04 \cite{Ostapchenko:2010vb}. The errors of the $\star$-ed parameters are shown in gray in Fig.~\ref{fig:zfactors}.

The fit can optimize the central values of parameters for which we have constraining power and explore a wide range of values for those without data constraints. We apply bounds to all the fit parameters to guarantee solutions with a physical meaning, although this is only used by the minimizer for the highest energy extrapolation parameters, where the data has little constraining power. Thanks to the inclusion of parameters up to 2~PeV, the error estimates on the lepton fluxes that our fit resturns are realistic, in the sense that they include all the constraints provided by the muon data, and only rely minimally on model assumptions.

We use the HESSE method implemented in i{\sc Minuit} to approximate the covariance matrix for all fitted parameters. Uncertainties on single parameters are taken from the square-root of diagonal elements but the full covariance matrix is used to compute uncertainties in the fluxes. The covariance matrix and best fit values are included in the code release accompanying this work.

The accuracy and robustness of the fitting procedure, as well as the precision expected on all parameters, was tested by injecting pseudo-measurements at the same energies reported by the experiments. An Asimov data set and ensembles of pseudo-data with statistical fluctuations were used to test that the fit recovers the injected values with minimial biases and to estimate the expected precision on each parameter. 

We assessed the magnitude of the uncertainties on the atmospheric density in our study by computing the maximal muon flux variations expected throughout the year for the location of each experiment, using NRLMSISE-00. We found these differences to be small, within 2\% from a globally averaged atmosphere, explained by the moderate latitudes at which the selected experiments were conducted. In our calculations we use a localized and seasonally adjusted flux for each experimental location, ensuring that the error caused by uncertainties on atmospheric description is kept well below the level of seasonal variations. Since the most precise data set used has uncertainties of at least 3\%, we neglect any potential error on our atmospheric modeling in our fit.

\subsection{Muon data fit results}

After performing the fit, we estimate the agreement between data and our model by computing a $\chi^2$ (using only the first term of Eq.~\ref{eq:modchi2}), and obtained a value of 199 with approximately 217 degrees of freedom. The number of degrees of freedom was estimated as number of data points minus the number of free parameters in the fit. This results in a a $p$-value of 81\%, indicating excellent agreement between the corrected data and the corrected model.

\begin{figure}
\centering
\includegraphics[width=0.72\textwidth]{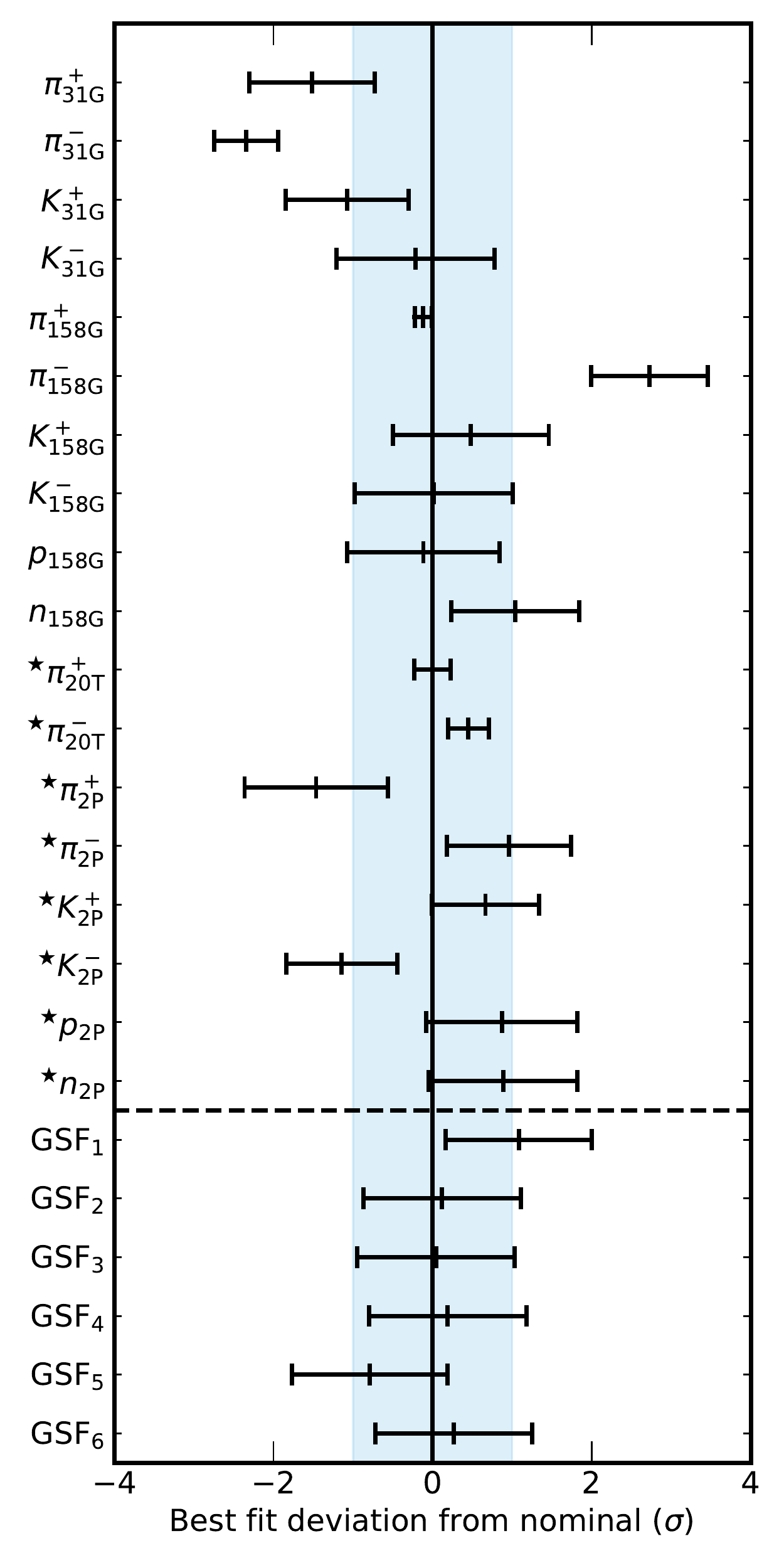}
\caption{Deviation between the best fit point and the nominal values of all parameters that modify the muon and neutrino flux, in terms of their uncertainty. The horizontal bars show the precision of the fit in units of the initial error.}
\label{fig:results}
\end{figure}

\begin{figure}
\centering
\includegraphics[width=0.9\textwidth]{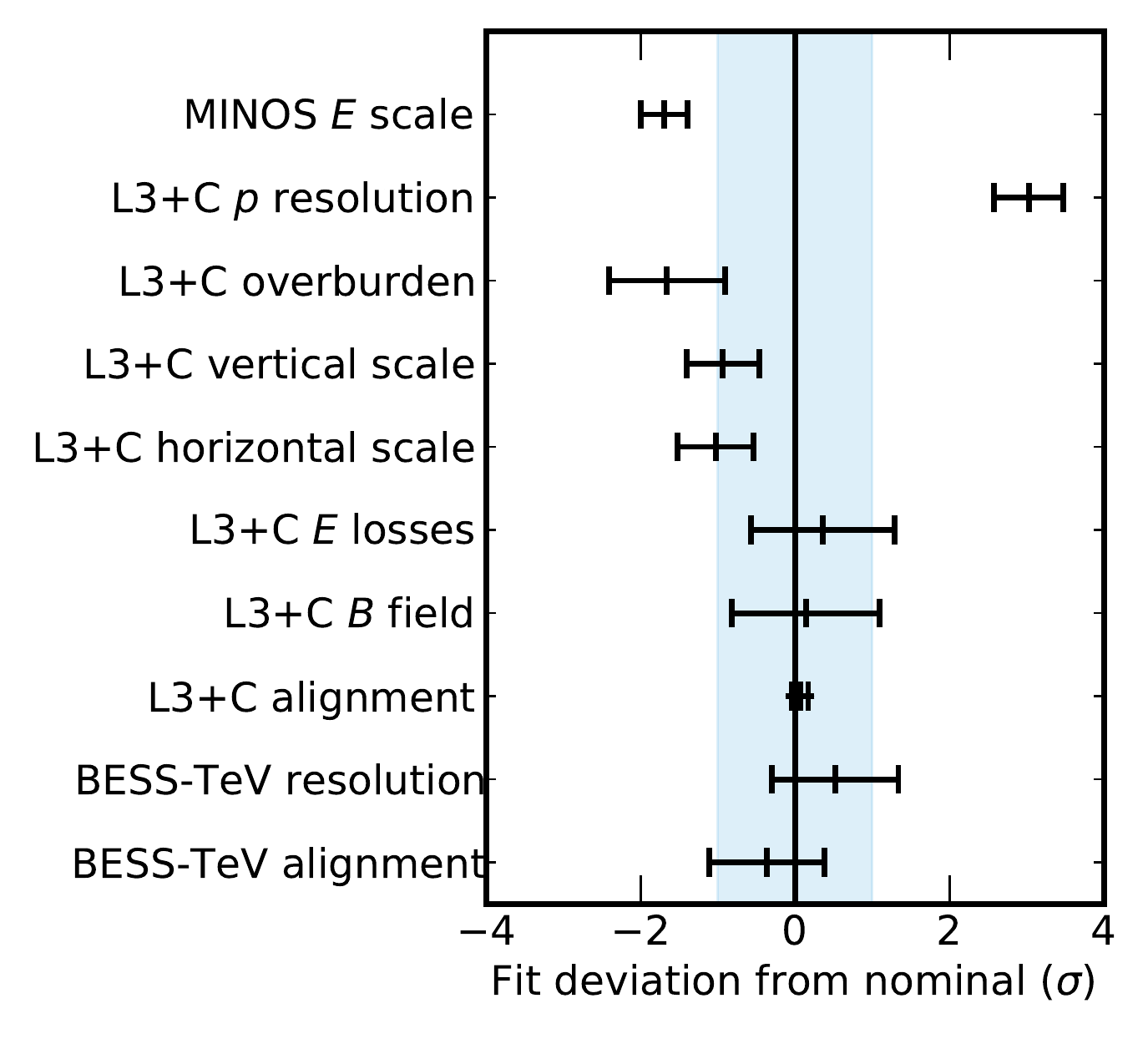}
\caption{Deviation between the best fit point and the nominal values of all parameters that modify the flux and charge ratio reported by the experiments used in the study, in terms of their uncertainty. The horizontal bars show the precision of the fit in units of the initial error.}
\label{fig:syst_results}
\end{figure}

\begin{figure*}[p]
\centering
\includegraphics[width=0.42\textwidth]{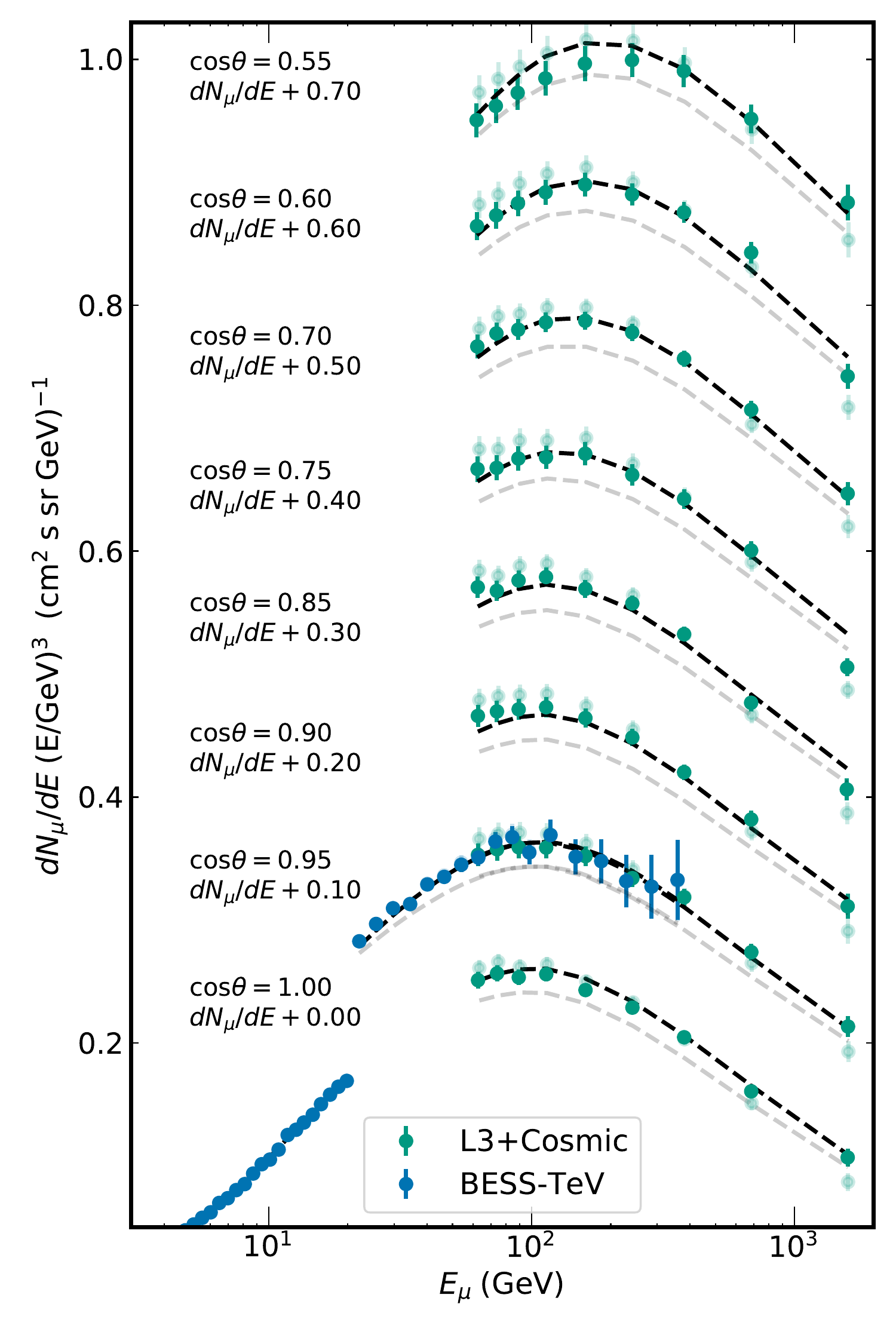}
\hspace{10.00mm}
\includegraphics[width=0.42\textwidth]{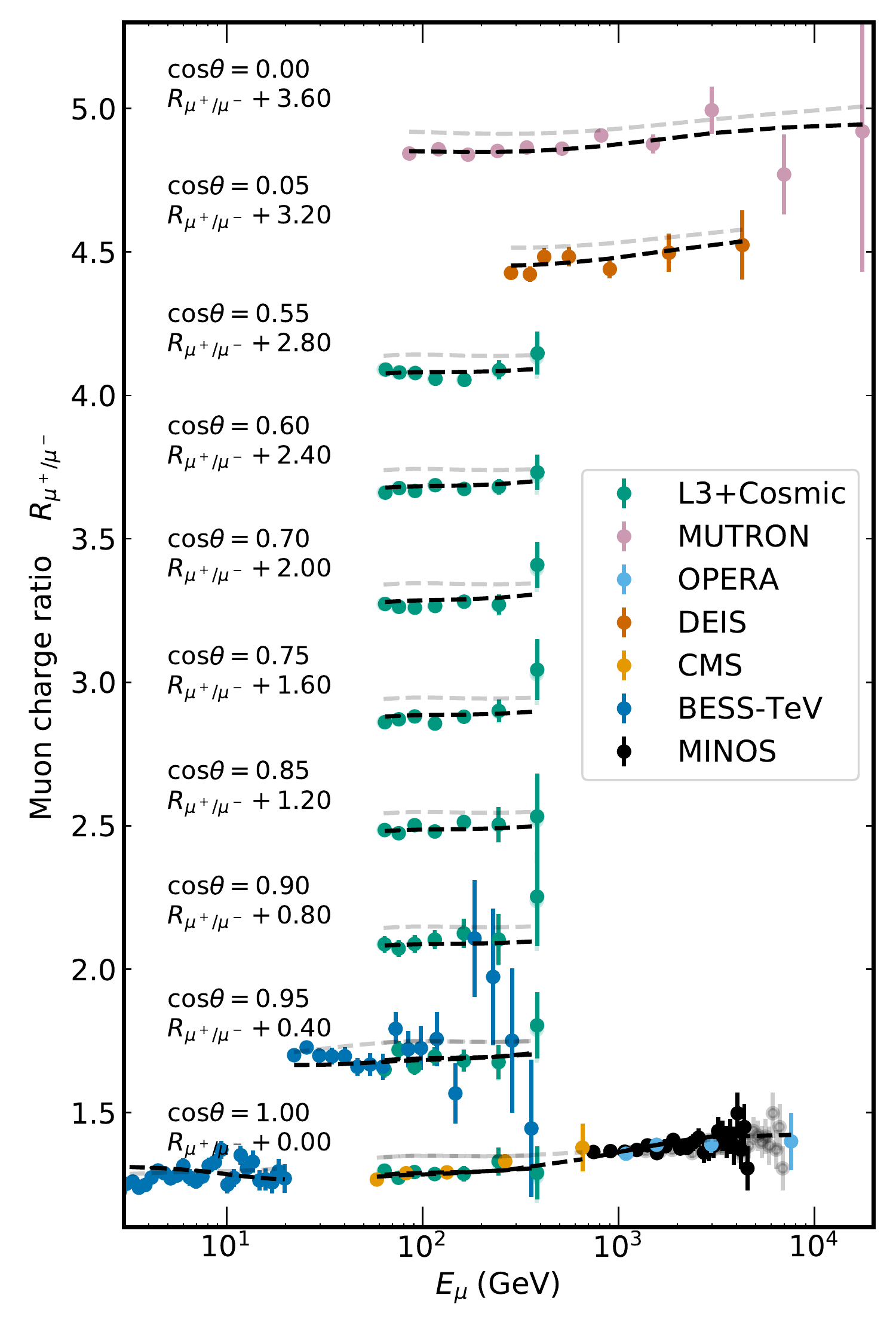}
\includegraphics[width=0.92\textwidth]{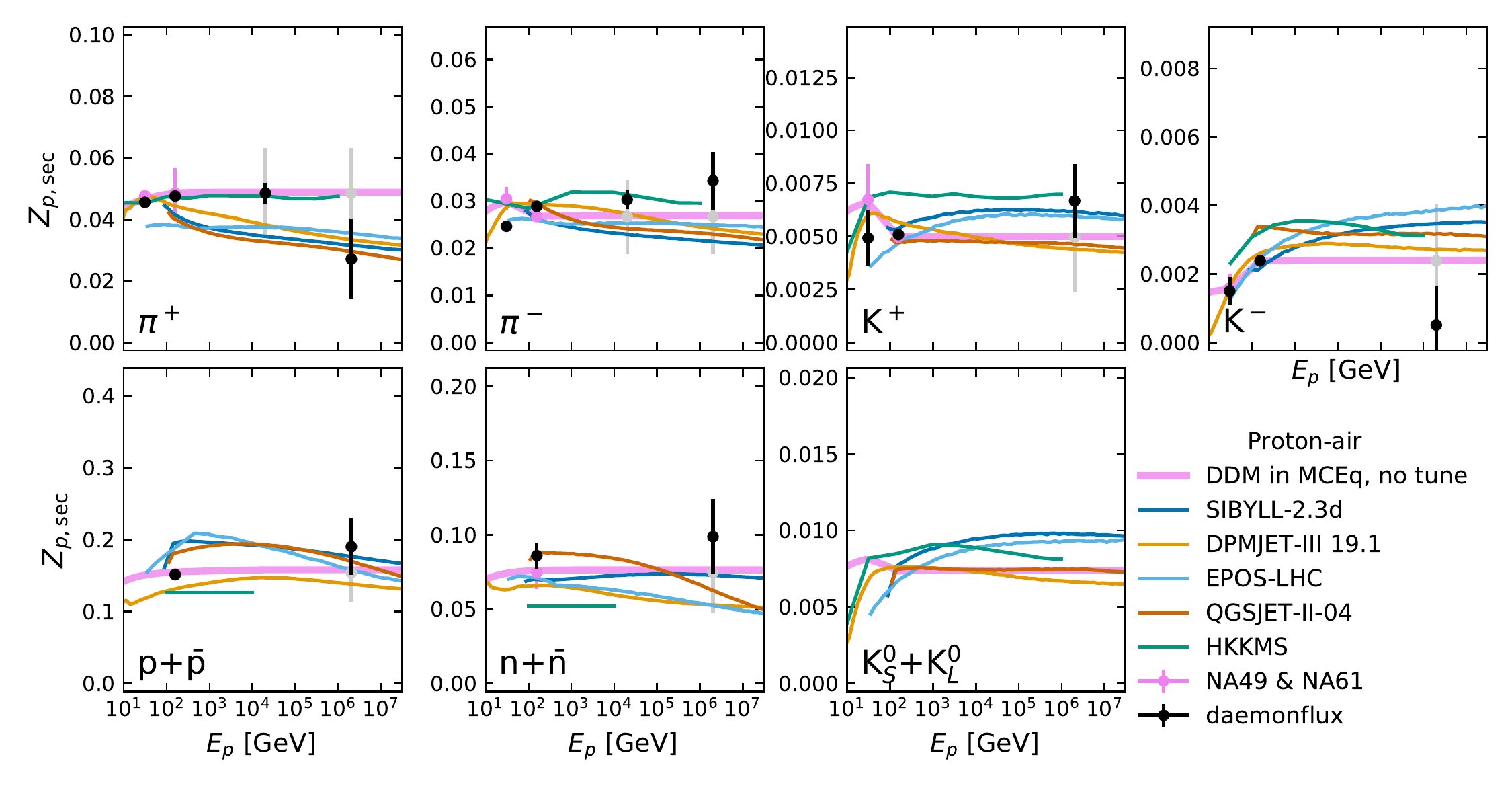}
\caption{
Muon fluxes for near vertical incoming directions (top left) and muon charge ratio for all data used in the fit (top right). Solid colors correspond to the experimental data and the flux calculation after the fit. The data and predictions prior to the fit are shown with a transparency. A factor has been added to fluxes from different incoming directions to be able to show them in the same figures.\\
Spectrum-weighted moments (Z-factors) shown as a function of energy for air target (bottom panel). They were computed assuming an incoming spectral index $\gamma=1.7$ following Eq.~\ref{eq:zfactor}. The result of this work is labeled \daemonflux{} and is shown in black. The grey points, located at the same energies (if visible) correspond to the starting value and uncertainty prior to the fit. Shown for comparison are the corresponding Z-factors from the DDM, HKKMS, DPMJet, Sibyll, QGSJET and EPOS-LHC \cite{Roesler:2000he,Fedynitch:2015kcn,Ostapchenko:2010vb,Pierog:2013ria}. Uncertainties for the $^\bigstar$-parameters are estimated from the spread of these models.
}
\label{fig:zfactors}
\end{figure*}

Figures~\ref{fig:results} and~\ref{fig:syst_results} show the best fit for all the fit parameters as deviations from their nominal expectation, in units of the initial error. The muon data introduces tight constraints to some of the pion yields. Most of the corrections are within the estimated initial error, and two of them show significant deviations. Note, however, than even in these cases the resulting particle yields fall within the range predicted by modern models, as shown in Fig.~\ref{fig:zfactors}, and the deviations appear strong due to the small errors reported by NA49 and NA61. Notably, the muon data is capable of providing meaningful constraints to two $^\bigstar$-ed parameters, $\vheIpip$ and $\vheIpim$. The fit modifies the yields of kaons, but their uncertainties are only slightly reduced. 

On the cosmic ray side, only the $\GSFI$ and $\GSFV$ parameters are changed from their initial value. For the energies that affect these studies, both parameters effectively act as spectral index corrections and have the same effect on both muons and neutrinos. On the experimental side, most corrections are within their uncertainties, with the exception of one L3+C parameter, which require strong pulls to explain the data. This behavior was observed already in the compatibility tests, where a similar pull was required to bring L3+C data with agreement with other experiment, and is therefore not a concern.

\begin{figure}
\centering
\includegraphics[width=0.95\columnwidth]{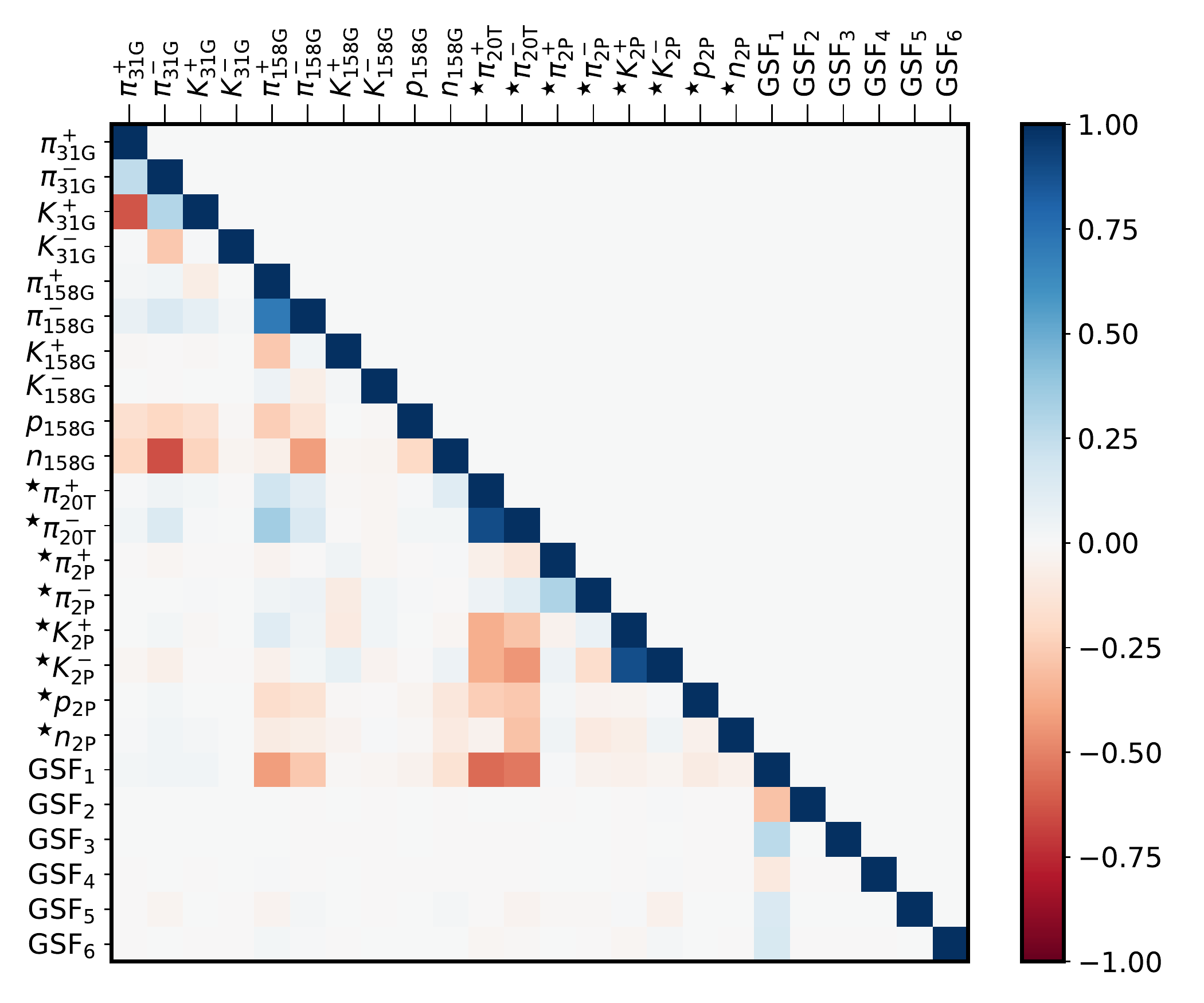}
\caption{Correlation between physics parameters obtained from the fit to the muon data.}
\label{fig:phys_corr}
\end{figure}

The correlations between the different parameters of interest are shown in Fig.~\ref{fig:phys_corr}. The hadronic yields at low energies show strong correlations, in particular between $\lepip{}$ and $\lepim{}$. At higher energies, the $^\bigstar$-ed parameters of \pions and \kaons also show negative correlations, possibly due to the \kaons being a sub-dominant contribution to the muon fluxes. $\GSFI$ is anticorrelated with \pions above its pivot point, since a change in a spectral index can be canceled by modifying these yields. Another implication of this is that the total error on the muon flux is lower than only the one from hadronic yields, as the fit cannot independently constrain hadronic and GSF parameters (see correlation matrix shown in Fig.~\ref{fig:phys_corr}).

The muon fluxes and muon charge ratios obtained from the analysis can be seen in Fig.~\ref{fig:zfactors}, which show the muon measurement data and the prediction from the model before and after the fit. The figure also shows the impact of the correction functions on experimental data. Prior to the fit, the prediction from GSF+DDM is in significant tension with the data, both for flux and charge ratio. After the fit is performed, and the data is corrected, the muon flux calculations can describe the data sets over the entire parameter space.



Figure~\ref{fig:zfactors} shows the results for hadronic yields as $Z$-factors, in comparison with predictions from multiple hadronic interaction models. Most of the resulting hadronic yields are near to one of the models tested, although none of the existing models follows the trend in energy of our result over all the yields. The assumption of scaling used in the DDM is significantly violated by the pion results, while for the kaons and baryons the error bars are large enough to be compatible with it. Table~\ref{table:zfactors} shows the same information, but in a numerical format. This can be compared with a similar table in~\cite{Fedynitch:2022vty}, where the $Z$-factors of the DDM are given.

\begin{table}
    \centering
    \begin{tabular}{llll}
\hline $p$C, 31 GeV& & & \\ \hline
$\pi+$  &  0.0455  &  $\pm$  &  2.5\%\\
$\pi-$  &  0.0246  &  $\pm$  &  4.1\%\\
$K+$  &  0.0049  &  $\pm$  &  26.5\%\\
$K-$  &  0.0015  &  $\pm$  &  27.6\%\\
\hline $p$C, 158 GeV& & & \\ \hline
$\pi+$  &  0.0476  &  $\pm$  &  1.8\%\\
$\pi-$  &  0.0289  &  $\pm$  &  2.0\%\\
$K+$  &  0.0051  &  $\pm$  &  5.0\%\\
$K-$  &  0.0024  &  $\pm$  &  3.4\%\\
$p$  &  0.1515  &  $\pm$  &  3.8\%\\
$n$  &  0.0860  &  $\pm$  &  10.2\%\\
\hline $p$C, 20 TeV& & & \\ \hline
$\pi+$  &  0.0485  &  $\pm$  &  6.9\%\\
$\pi-$  &  0.0303  &  $\pm$  &  6.7\%\\
\hline $p$C, 2 PeV& & & \\ \hline
$\pi+$  &  0.0271  &  $\pm$  &  48.6\%\\
$\pi-$  &  0.0343  &  $\pm$  &  18.0\%\\
$K+$  &  0.0067  &  $\pm$  &  26.0\%\\
$K-$  &  0.0005  &  $\pm$  &  227.8\%\\
$p$  &  0.1904  &  $\pm$  &  20.6\%\\
$n$  &  0.0988  &  $\pm$  &  25.8\%\\
        \hline
    \end{tabular}   
    \caption{Table of spectrum-weighted moments for $p$C resulting from the fit to muon data, assuming a spectral index $\gamma$=2.7.}
    \label{table:zfactors}
\end{table}

\subsection{Calibrated atmospheric neutrino fluxes}
\begin{figure}
\centering
\includegraphics[width=0.84\textwidth]{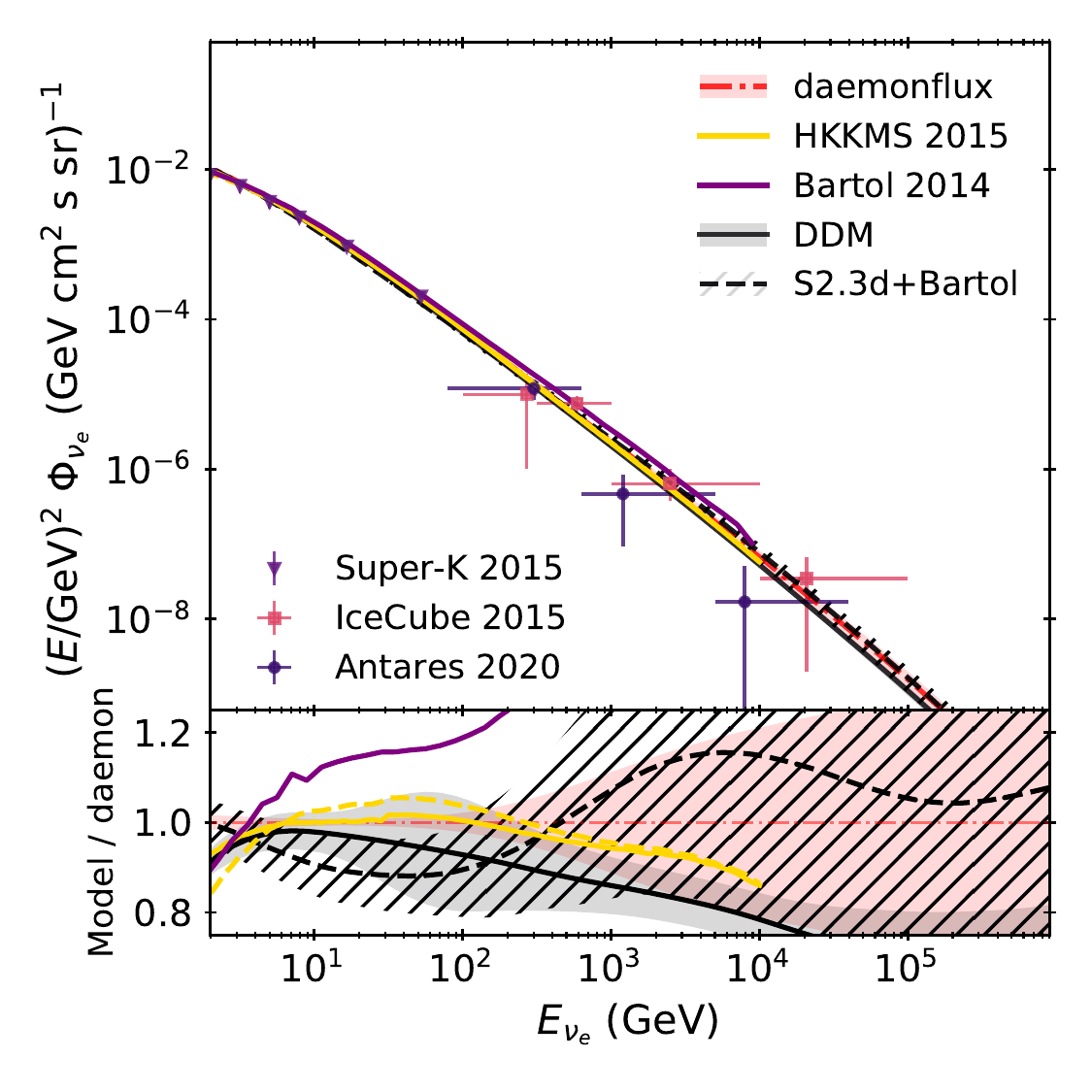}
\hspace{5.00mm}
\includegraphics[width=0.84\textwidth]{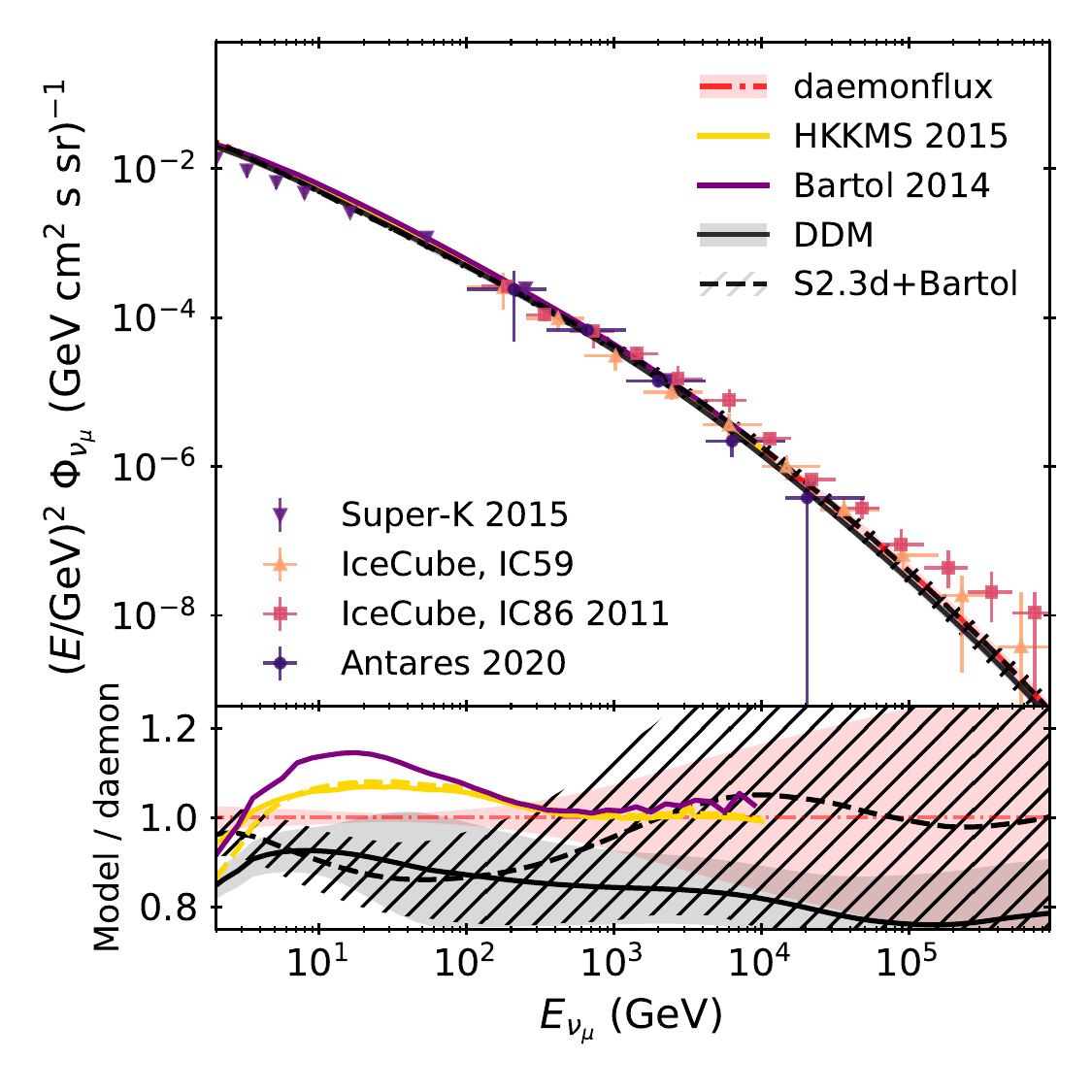} 
\caption{Conventional atmospheric electron neutrino (top) and muon neutrino (bottom) fluxes averaged over zenith angles shown together with data from Super-Kamiokande, IceCube, and ANTARES \cite{ANTARES:2021cwc,Richard:2015aua,Aartsen:2014qna,Aartsen:2015xup}. }
\label{fig:nuflux}
\end{figure}

The atmospheric neutrino fluxes after muon calibration are displayed in Fig~\ref{fig:nuflux}, along with predictions from various models and experimental data. The uncertainties for models computed with \mceq{} are shown as bands around the central prediction. The panel below the figures provides a comparison of each model to our calibrated result. 

Our result displays interesting differences from the HKKMS 2015 model, which has been used to interpret atmospheric neutrino data below 1~TeV by various experiments (see e.g.~\cite{Super-Kamiokande:2019gzr,IceCubeCollaboration:2021euf}). 
For $\nu_e$, our calculation predicts a flux about 5\% lower between~$\sim\SIrange[range-phrase=-,range-units=single]{40}{1000}{GeV}$, with the difference growing with energy. For $\nu_\mu$ the discrepancy between both models is closer to 8\% between~$\sim\SIrange[range-phrase=-,range-units=single]{5}{100}{GeV}$, then agreeing for higher energies. 

\begin{figure}
\centering
\includegraphics[width=0.95\textwidth]{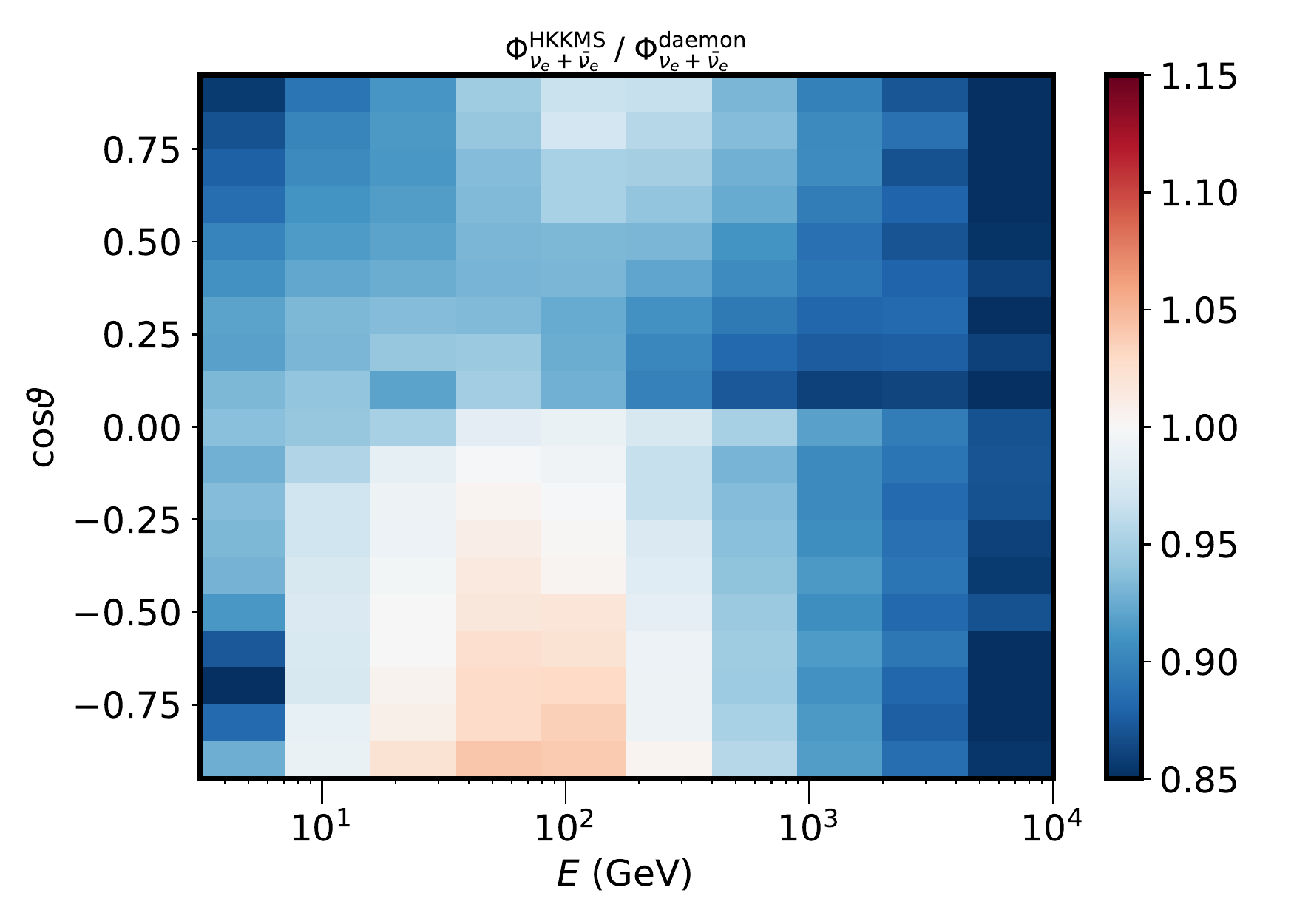}
\includegraphics[width=0.95\textwidth]{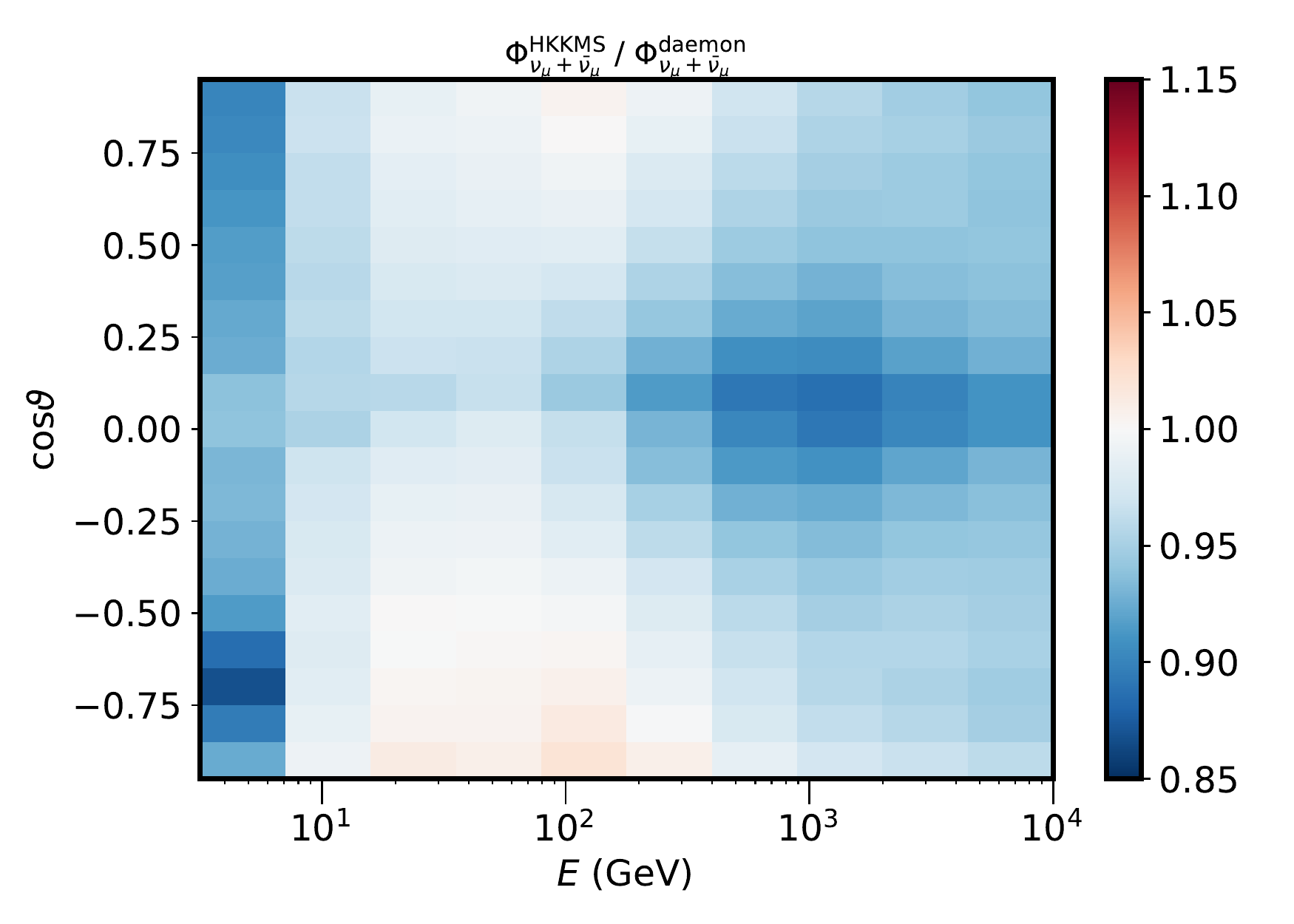}

\caption{Ratio of flux of electron and muon neutrinos and antineutrinos from the prediction from Honda et al and our calculation. The fluxes were scaled to the same value at $E_\nu = 25$~GeV and $\cos\vartheta$=-0.5 to highlight their shape difference.}
\label{fig:numuflux_2d}
\end{figure}

The neutrino flux data has large uncertainties and does not allow for a clear differentiation between the models in the comparison. It should be noted that an alternative data-driven method previously predicted a lower $\nu_\mu$ flux, estimating neutrino fluxes from neutrino data but with significant uncertainties \cite{Gonzalez-Garcia:2006xta}. A more detailed display of these differences is shown in Fig.~\ref{fig:numuflux_2d}, where the ratio between our result and HKKMS 2015 is presented as function of neutrino energy and incoming direction. In this comparison one can see that our result predicts a zenith dependence different from that in HKKMS 2015.

\begin{figure}
\centering
\includegraphics[width=0.85\textwidth]{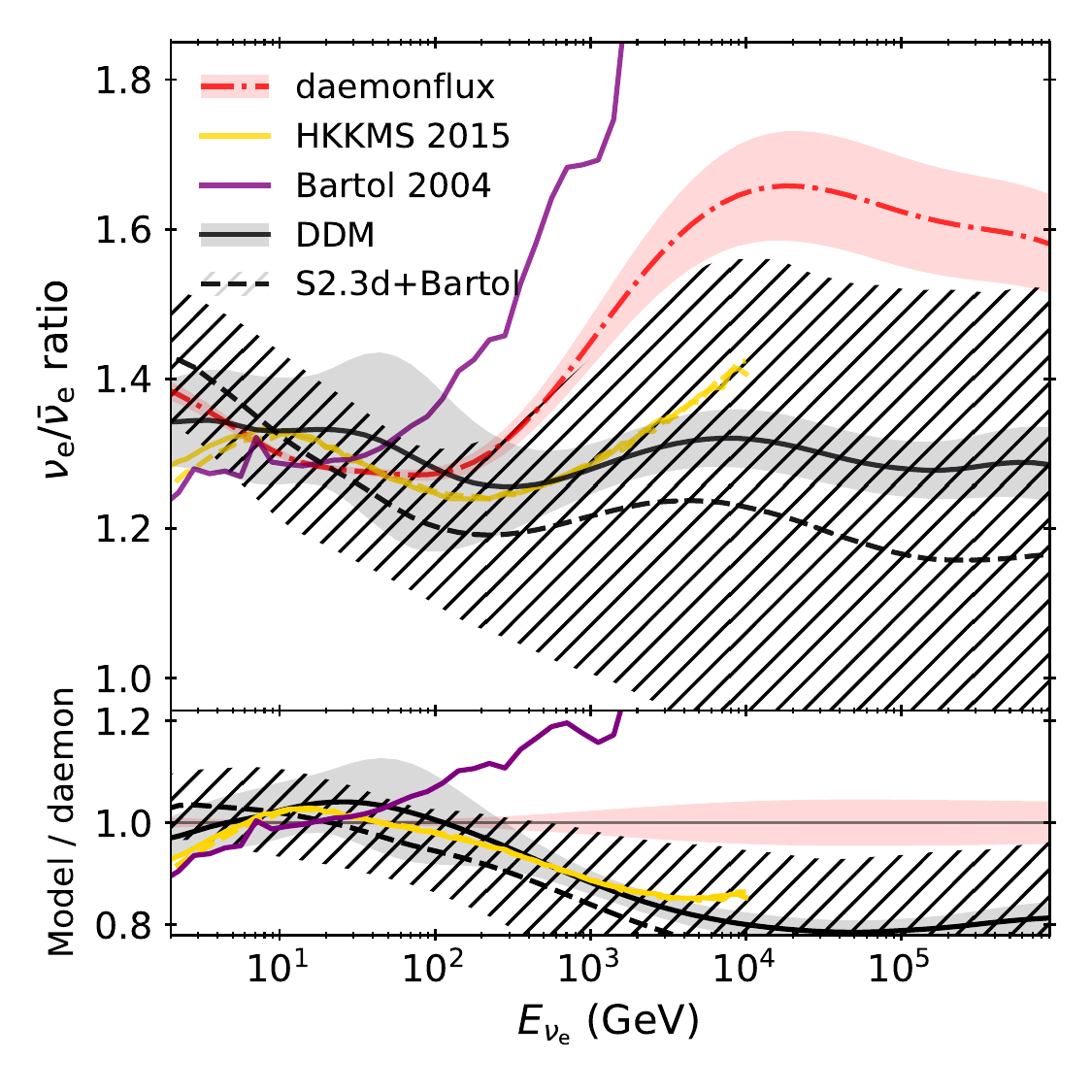}
\hspace{10.00mm}
\includegraphics[width=0.85\textwidth]{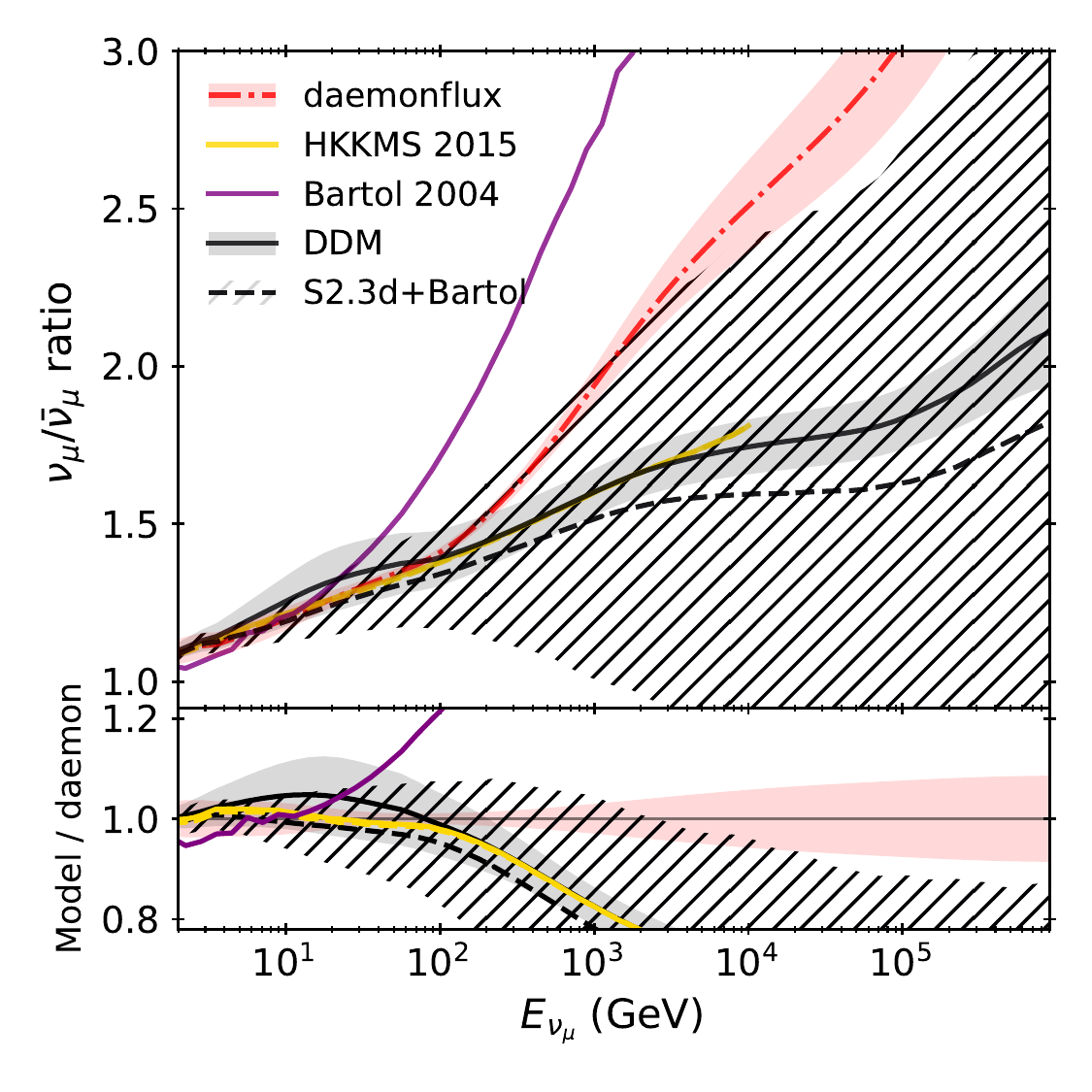} 
\caption{Neutrino-antineutrino ratios for the conventional, zenith-averaged atmospheric neutrino flux. The uncertainties are computed as described in~\cite{Fedynitch:2022vty}.}
\label{fig:nuratio}
\end{figure}

\begin{figure}
\centering
\includegraphics[width=0.85\textwidth]{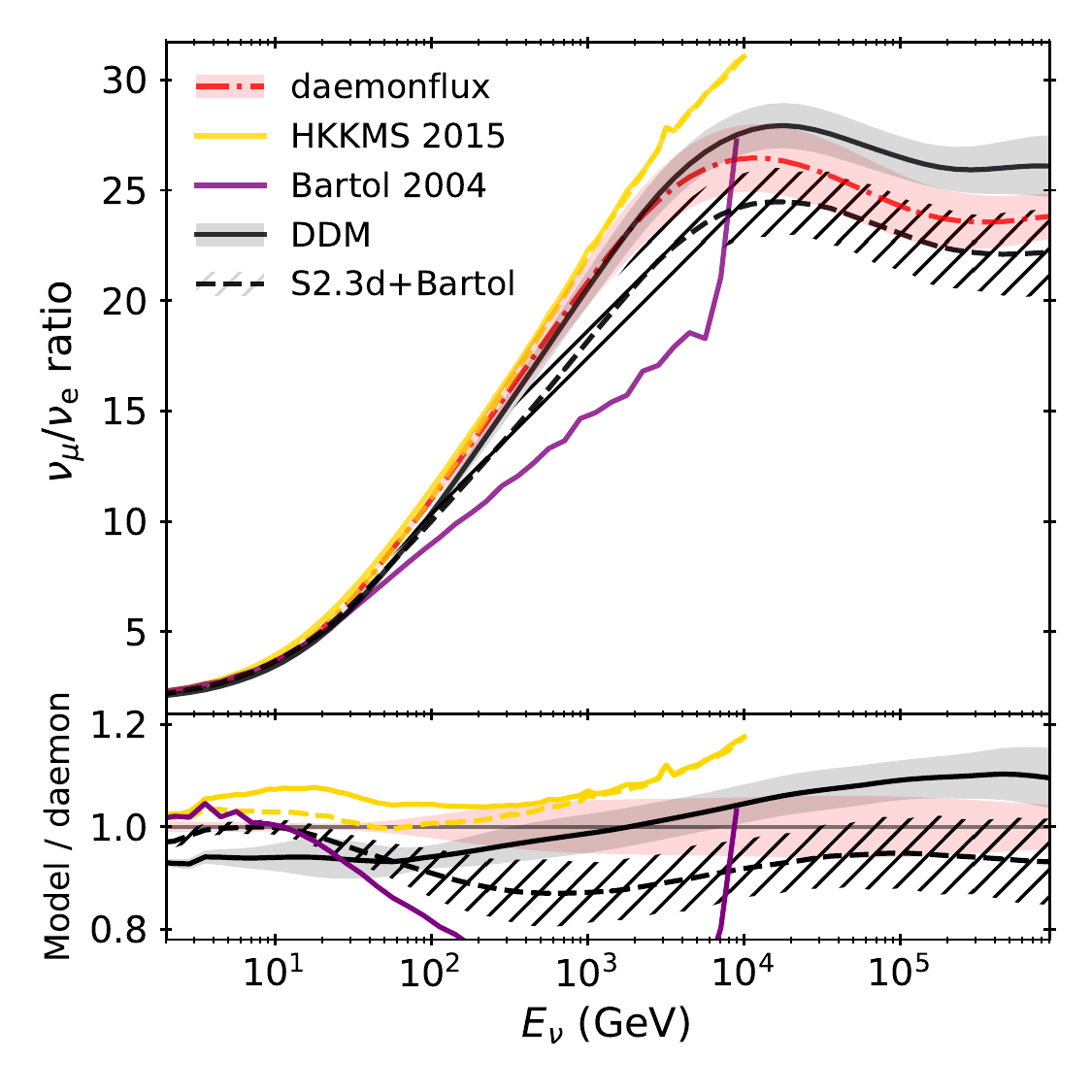} 
\caption{Ratio of the zenith-averaged conventional atmospheric fluxes of muon neutrinos to electron neutrinos as function of energy. The prediction from \daemonflux{} is about 10\% lower than the commonly used HKKMS 2015.}
\label{fig:flavorratio}
\end{figure}

\begin{figure}
\centering
\includegraphics[width=0.95\textwidth]{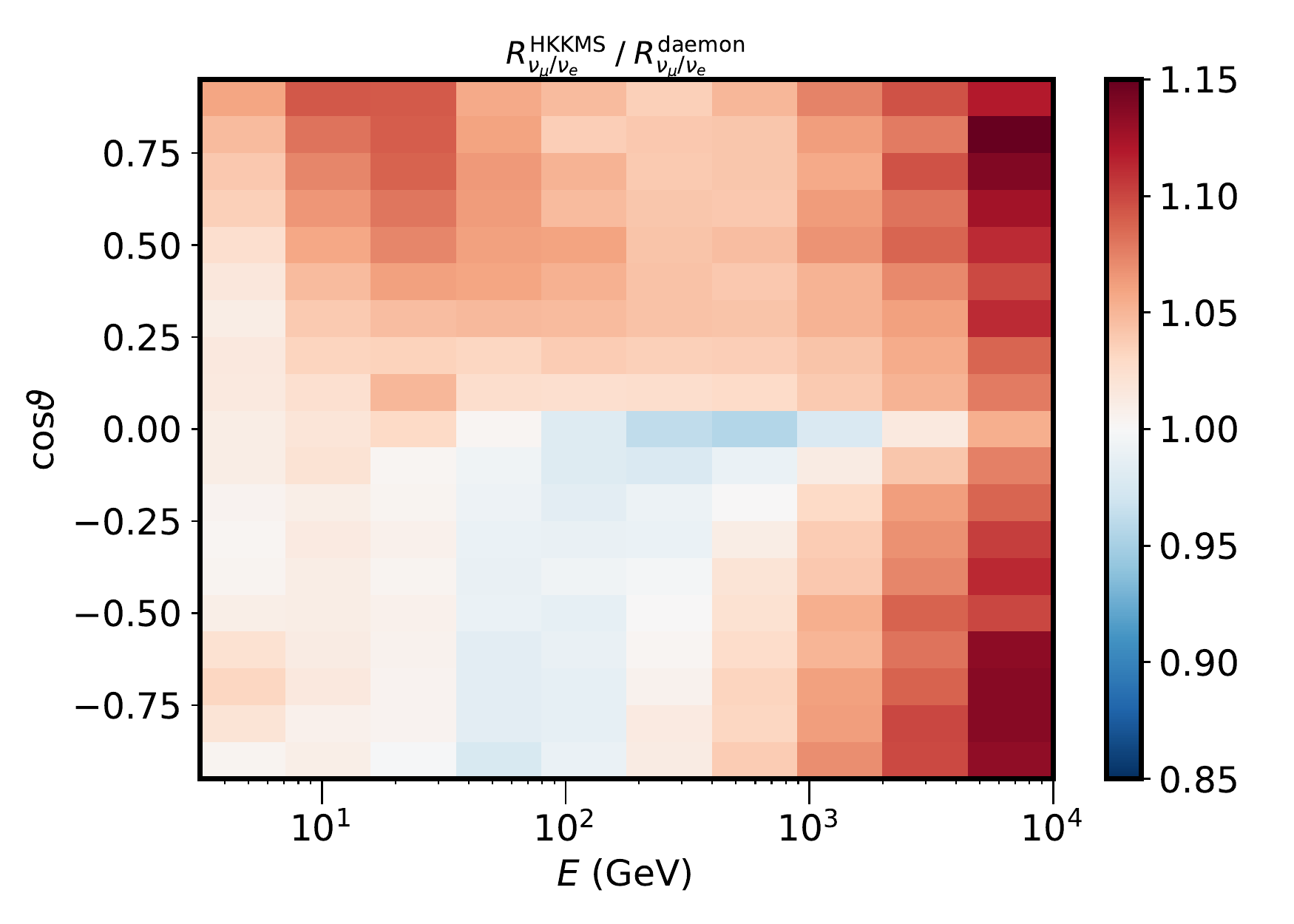}
\caption{Flavor ratio $\Phi_{\nu_\mu} / \Phi_{\nu_e}$ predicted by Honda et al divided by our results. The fluxes include neutrinos and antineutrinos. Both ratios were scaled to the same value at $E_\nu = 25$~GeV and $\cos\vartheta$=-0.5 to highlight their shape difference.}
\label{fig:flavorratio_2d}
\end{figure}

Neutrino to antineutrino ratios are shown in Fig.~\ref{fig:nuratio}. Calculations agree below 100~GeV, but start diverging after that point. Our result predicts a ratio that is higher than commonly used models, although still lower than the Bartol~2004 calculation. The ratio of muon to electron neutrinos, depicted in Fig.~\ref{fig:flavorratio}, shows differences of the order of 5\% with the HKKMS 2015 model, and moderate agreement with DDM and \sibyll{}. This ratio is one of the key handles to measure atmospheric neutrino oscillations, since the electron neutrino flux is expected to be largely unaffected while the muon neutrinos can oscillate into the tau flavor. A 2D comparison of this ratio between our result and HKKMS 2015 as a function of neutrino energy and incoming direction is shown in Fig.~\ref{fig:flavorratio_2d}. Interestingly, the ratio of both models is fairly close to one below 100~GeV and for $\cos\theta<0$, where atmospheric neutrino oscillations are expected.

\begin{figure*}
\centering
\includegraphics[width=0.95\textwidth]{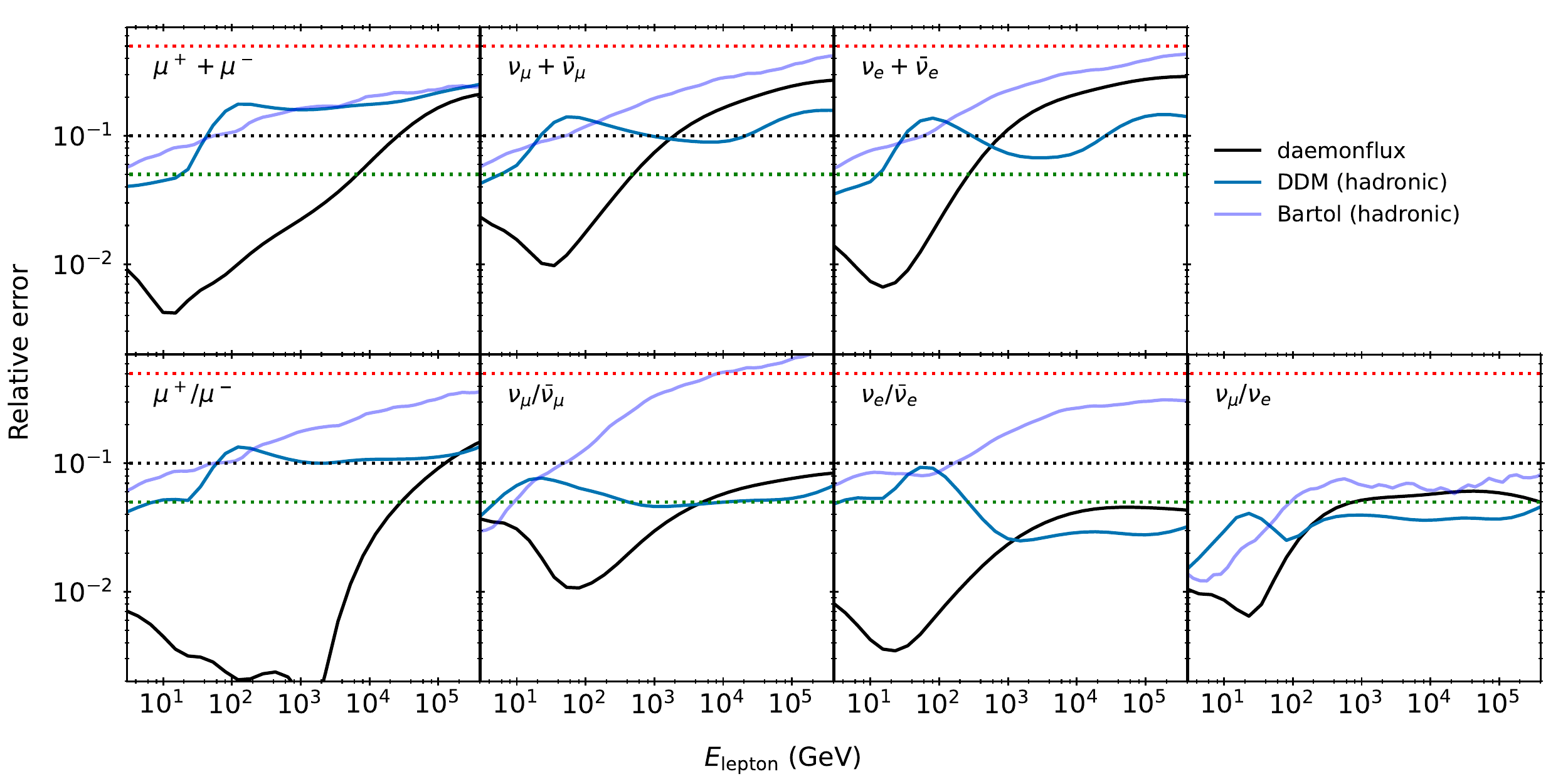}
\caption{Comparison of errors in different schemes. The original DDM underestimates tends to underestimate errors at high energies with its simple Feynman-scaling assumption for energy extrapolation. The \daemonflux{} reduces errors at lower energies, where muon data constraints are stronger. However, the errors are larger at high energies due to missing constraints from surface muon data. The MCEq implementation of the error estimate from Ref.~\cite{Barr:2006it} (Bartol) is also shown for comparison.}
\label{fig:nu_unc}
\end{figure*}

The uncertainties in the fluxes and ratios of muons are under 10\% up to energies of 10 TeV. The ratio is particularly well constrained. For neutrinos the error stays below 10\% up to 1 TeV and increase at higher energies to $\sim30\%$. The errors on the neutrino ratios are constrained to below 10\% over entire energy range of this work. Fig.~\ref{fig:nu_unc} displays the uncertainties for all the channels discussed, compared to the uncalibrated DDM model and a calculation using the \sibyll{-2.3d} model with an uncertainty estimate from Ref.\cite{Barr:2006it}.

\section{Discussion and outlook}

\subsection{Choice of DDM extrapolation parameters}

In the DDM model, the standard evolution of the $Z$-factors results in a rigid model at high energies. To better fit the data, we needed to introduce additional freedom. However, determining the number of parameters and their exact locations was not a straightforward task. We tested combinations that included up to three extra points per hadronic yield, distributed between TeV and PeV projectile energies. To avoid correlations between these parameters, we reduced the number to match the precision of the available data, and evaluated the effect of each parameter on the fluxes (as shown in Fig.~\ref{fig:impact_hadrons}) to avoid overlap between two non-adjacent points. The error on these parameters was conservatively estimated from the range predicted by various models. This choice particularly affects the highest energy kaons, where the data has the weakest constraints. Although adding a prior to these parameters was not ideal, it was necessary to prevent overfitting of statistical fluctuations and to keep the parameters within a physically meaningful range. To validate our results, we compared the outcome of a fit with and without priors on the $^\bigstar$-parameters, and found them to be in agreement within their uncertainties.

\subsection{Relevance of individual data sets}

\begin{figure}
\centering
\includegraphics[width=0.75\textwidth]{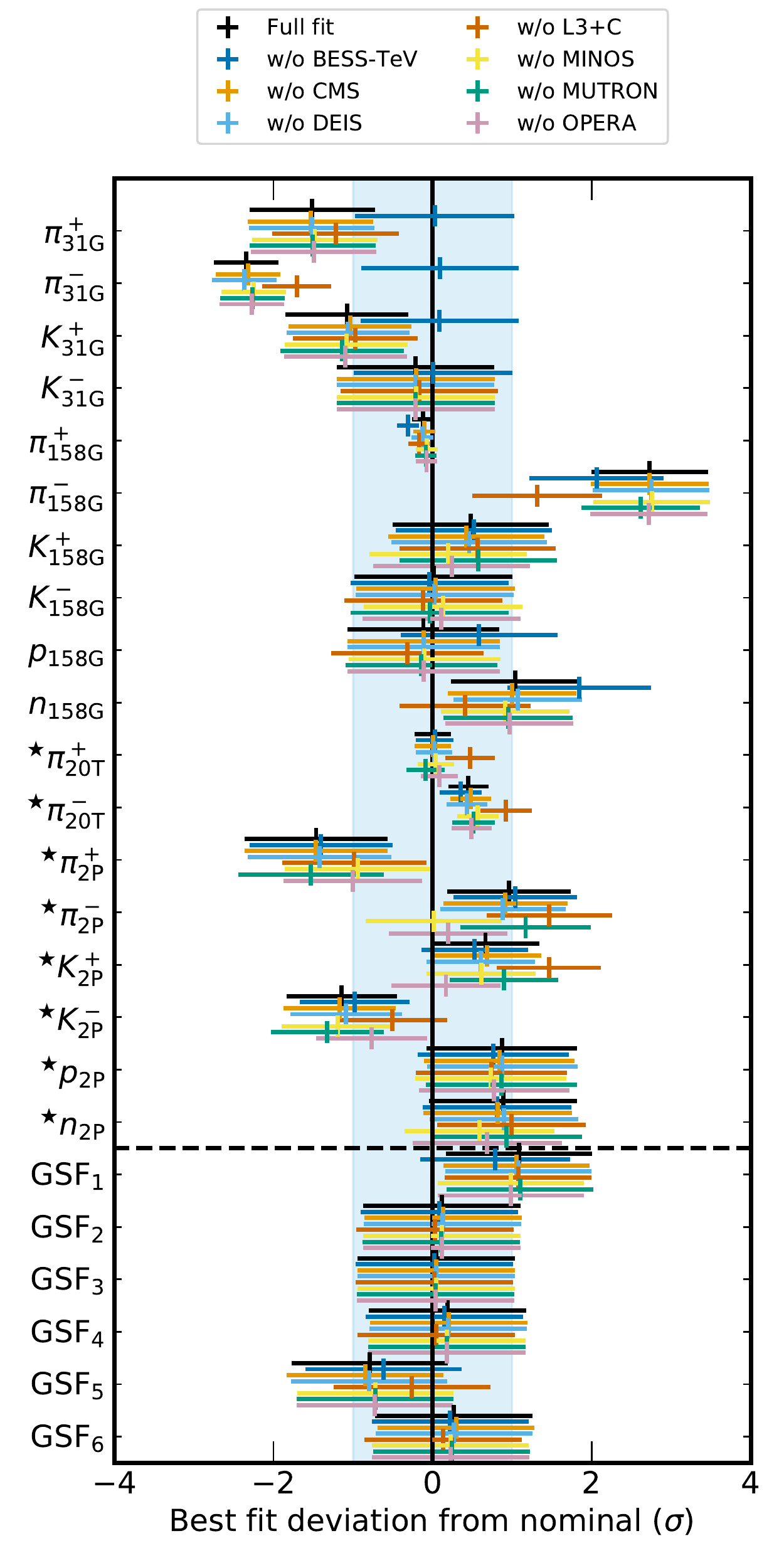}
\caption{Comparison between the best fit point of the flux using all the data sets and the fits obtained when removing one data set at a time. The format of the image is the same as Fig.~\ref{fig:results}.}
\label{fig:results_nminus1}
\end{figure}

The data selection criteria for our work were stringent, leading to the inclusion of only seven experiments in the final fit. To evaluate the contribution of each experiment, we performed a systematic analysis by removing each experiment one by one and re-fitting the data. The results are shown in Fig.~\ref{fig:results_nminus1}.

In the full data set, the vertical direction and medium energy range (80~GeV to a few TeV) are covered by multiple experiments, but the low energy regime is only represented by BESS-TeV. This is reflected in the results of the systematic analysis. Removing BESS brings the low energy corrections to their original value and error, as the fit loses its constraining power. The relevance of L3+C data can be seen in the pion and kaon parameters at 158~GeV and above, where corrections of the order of the precision of the fit are needed to explain the data. The significant change in the $\nu/\bar{\nu}$ ratios is a result of the charge ratio measurements from OPERA and MINOS. The removal of either of them does not change the conclusion. 

\subsection{Underground data}
Measuring the atmospheric muon spectrum above a few TeV in spectrometers is challenging due to the low flux and the strong magnetic fields required to determine the momentum. This results in significant statistical uncertainties in this energy regime, which reduces the precision of the calibration. In addition, the neutrino flux in this regime is dominated by kaons, which are a secondary component for the muon flux. To mitigate these challenges, data from large volume experiments located deep underground can be included. These experiments measure the integrated flux above a given energy, defined by the amount of material present in a specific direction. Recent studies have shown that the uncalibrated GSF+DDM model can reasonably explain the data~\cite{Fedynitch:2021ima}. Thus, it may be possible to incorporate these results into future updates of this work.

\section{Conclusion}

The \daemonflux{} atmospheric lepton flux model represents a significant departure from previous methods of calculating fluxes and characterizing their uncertainties. By incorporating data-driven models for the cosmic ray spectrum (GSF6) and secondary particle yields (DDM) into an accurate cascade equation solver (MCEq), we can propagate the uncertainties of these models to the prediction of atmospheric muon fluxes, which are closely related to atmospheric neutrino fluxes. Using precise spectrometer data of muon fluxes at the surface, we developed a fitting framework that uses corrections to the initial model uncertainties as fit parameters to produce a best fit and a fully data-driven estimate of uncertainties. The result is a flux model with the lowest uncertainties to date, reducing them by dozens of percent in energy ranges where relevant muon data is available. This energy range covers the signal ranges for atmospheric neutrino oscillation for standard and sterile neutrino scenarios at neutrino telescopes.
To minimize biases in the parameterization at high energies, we introduced more flexibility in the model at energies where it must extrapolate beyond the current limits of fixed-target experiments.

In future iterations, the precision and uncertainty at high energies can be improved by including existing underground muon measurements and data from neutrino telescopes like IceCube, Baikal, and KM3NeT. To further improve our understanding of the full parameter space, measurements from ongoing experiments that could report horizontal muon fluxes and muon fluxes at higher energies would be very valuable.


 Although our technique could be applied at energies below 5 GeV, the signal region of Super-Kamiokande and Hyper-Kamiokande, the modeling of flux and experimental uncertainties requires extraordinary detail of the geomagnetic, atmospheric, and solar conditions at the experimental site used for calibration, as well as significant effort in understanding the systematic uncertainties and corrections applied by that particular experiment.

\acknowledgements
This research was undertaken thanks in part to funding from the Canada First Research Excellence Fund through the Arthur B. McDonald Canadian Astroparticle Physics Research Institute, WestGrid and Compute Canada (www.computecanada.ca). The author, A.F., is grateful for the supportive environment provided by Prof.~Hiroyuki Sagawa's group at the ICRR, where he was able to make substantial progress on this work as a recipient of the JSPS International Research Fellowship (JSPS KAKENHI Grant Number 19F19750). The authors acknowledge the invaluable computational resources and support provided by the Academia Sinica Grid-Computing Center (ASGC), which is supported by Academia Sinica. The authors also extend their thanks to Maria Liubarska for testing the preliminary output of this work, Michael Unger for his valuable insights and feedback on the interpretation of L3+C data, and to the ChatGPT AI for contributing to linguistic improvements throughout the text.

\bibliography{references}

\appendix
\section{Impact of DEIS flux data\label{appendix:deis}}



\begin{figure*}
\centering
\includegraphics[width=0.35\textwidth]{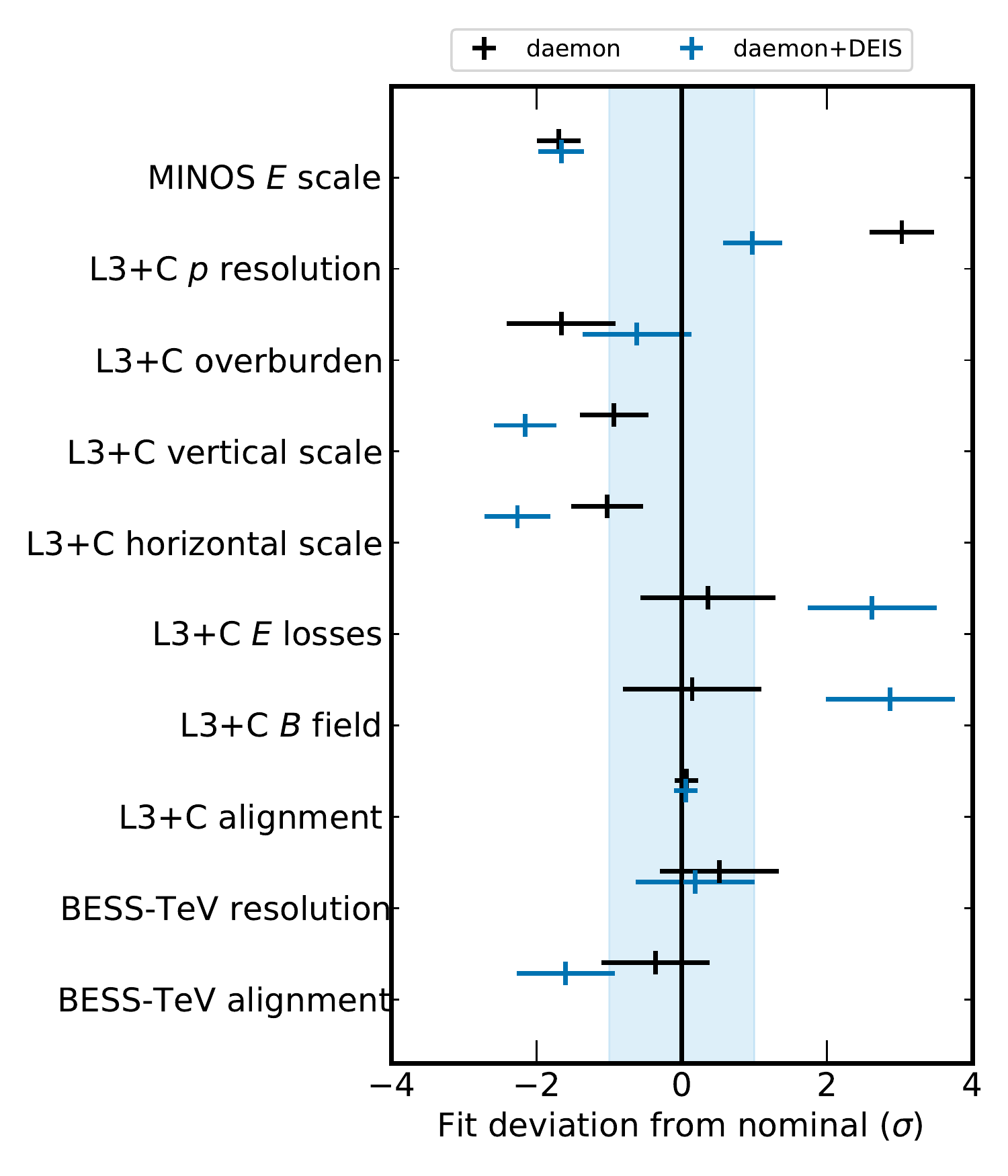}
\includegraphics[width=0.45\textwidth]{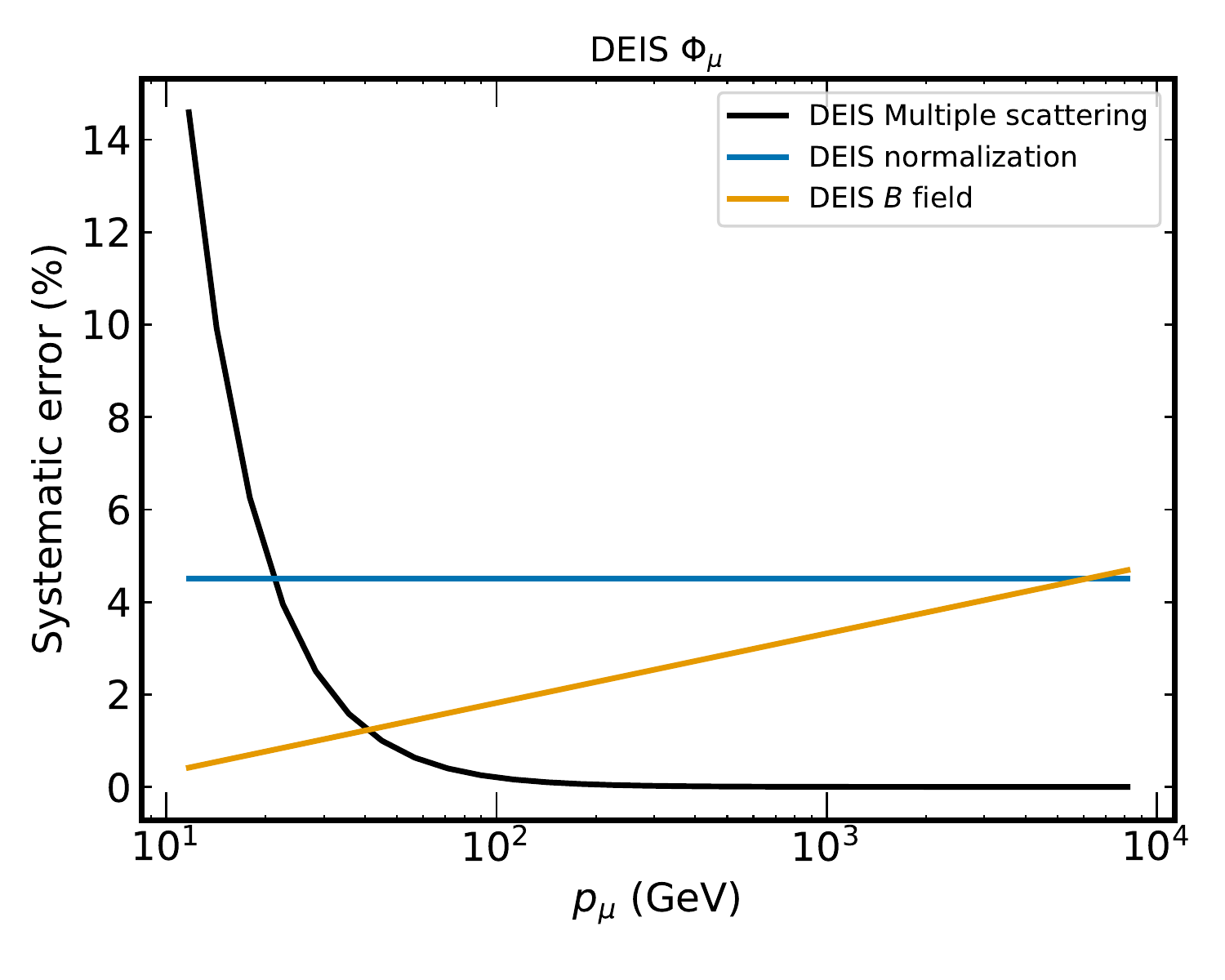}
\caption{(left) Impact of including the DEIS muon flux measurements on the best fit values of the experimental corrections of the fit. The format of the image is the same as Fig.~\ref{fig:results}. (right) Parametrized errors on the fluxes reported by DEIS, estimated from the information given in~\cite{Allkofer:1985ey}. These errors are assumed to be uncorrelated bin-by-bin, and are thus summed to the statistical ones in quadrature.\label{fig:deis_syst}}
\end{figure*}

Near-horizontal muon data can be very useful in constraining the different parameters of the fit, as illustrated in Fig.~\ref{fig:impact_zenith}. With this in mind we tested the DEIS data set reported in~\cite{Allkofer:1985ey}, where the following systematic errors are discussed:
\begin{itemize}
    \item A magnetic field uncertainty that produces momentum dependent errors in the spectrum from 1\% at $p_\mu=$50~GeV to 3\% at $p_\mu=$1~TeV. This was parameterized as a linear dependence on $\log_{10}(p)$.
    \item A total normalization error of 4.5\% correlated for all energies.
    \item Multiple scattering and energy losses that impact low momenta data, from 20\% at 10~GeV to 1\% at 45~GeV. This effect was also parameterized as a linear function on $\log_{10}(p)$.
\end{itemize}
The parameterized impact of these errors is shown in Fig.~\ref{fig:deis_syst}. 

\begin{figure}[!b]
\centering
\includegraphics[width=0.72\textwidth]{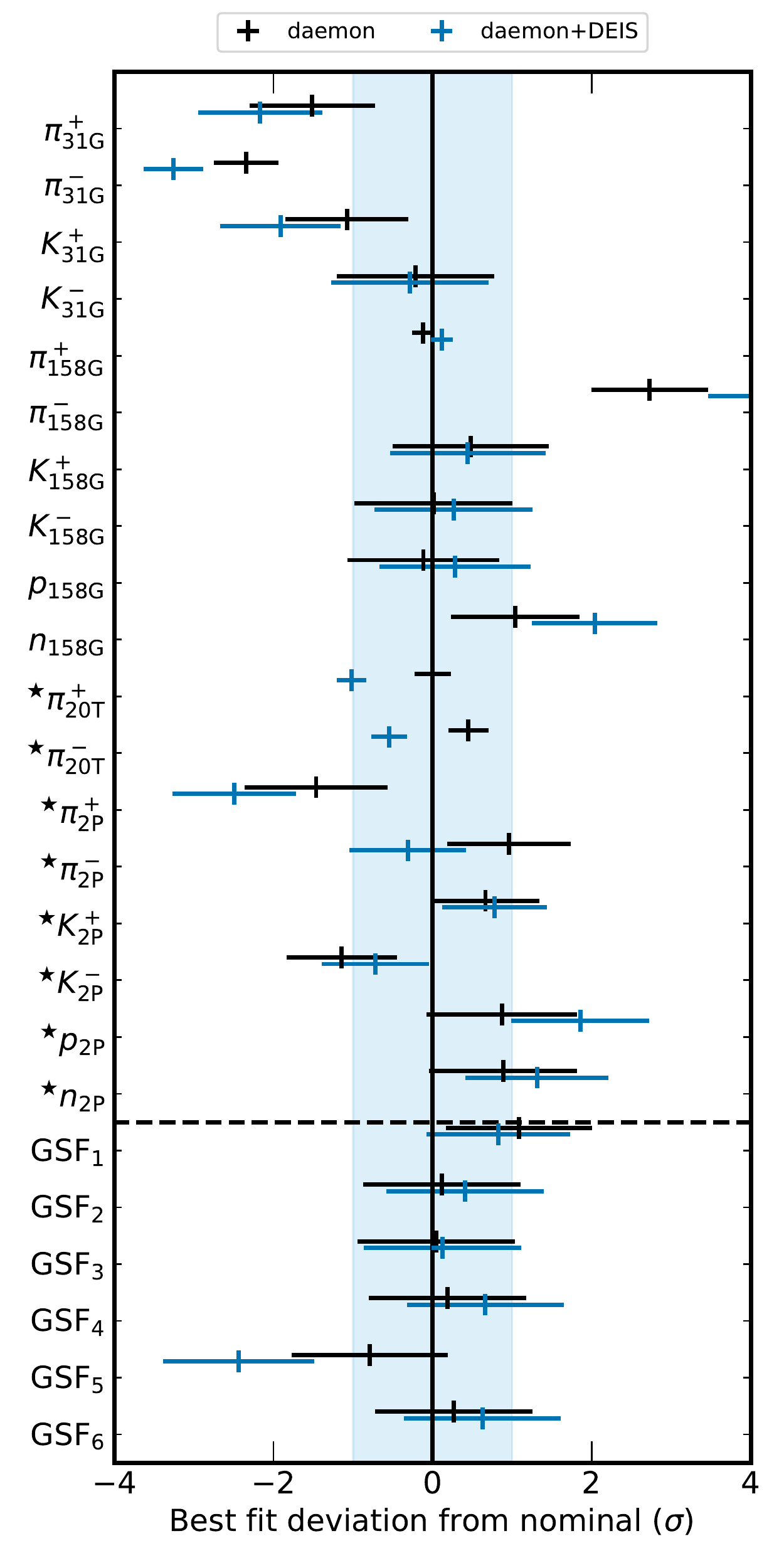}
\caption{Impact of including the DEIS muon flux measurements on the best fit values of the parameters of interest. The format of the image is the same as Fig.~\ref{fig:results}.}
\label{fig:results_deis}
\end{figure}

We carried out the smoothness test on these data, but found the low energy measurements to have a spread inconsistent with random statistical fluctuations. Applying the parameterized uncertainties as correction functions did not resolve the problem. Instead, these uncertainties were introduced by summing them in quadrature with the statistical errors, which produced a flux compatible with the smooth expectation. Using this approach, we were able to obtain a good fit to the data using the free parameters of the model. However, this result turned out to introduce significant pulls on multiple high energy parameters as shown in Fig.~\ref{fig:results_deis}. Due to DEIS, the $\vheIpip$ and $\vheIpim$ parameters are fit at a significantly higher value and $\lepim$ and $\hepim$ result in significantly larger pulls than those obtained without this data set.

The impact on the systematic parameters is given in the left panel of Fig.~\ref{fig:deis_syst}. To bring DEIS and L3+C into agreement, most L3+C correction functions need to be pulled well over their error estimate. Moreover, preliminary comparisons to underground muon measurements also indicate a tension with the DEIS data. For these reasons, our final fit does not include the DEIS muon flux data. Nonetheless, the flux that results after including DEIS is provided as an option in the \daemonflux{} tool.



\end{document}